\documentclass[prb,twocolumn,showpacs,amsmath,amssymb]{revtex4-2}
  \usepackage{graphicx}
  \usepackage{color}
  \usepackage{bm,upgreek}

  \usepackage{fancyhdr}
  \usepackage[short,nodayofweek,level,12hr]{datetime}

\begin{document}
  \title{Onset of superconductor-insulator transition in an ultrathin NbN film under in-plane magnetic field studied by terahertz spectroscopy}

  \author{M. \v{S}indler$^1$}
\email{sindler@fzu.cz}
  \author{F.~Kadlec$^1$}
  \author{C.~Kadlec$^1$}

\affiliation{$^1$Institute of Physics, Czech Academy of Sciences, Na Slovance 2, 182 21 Prague 8, Czech Republic}

  \date{\today}




  \begin{abstract}
  Optical conductivity of a moderately disordered
  superconducting NbN film was investigated
  by terahertz time-domain spectroscopy in external magnetic field applied along
  the film plane.
   The film thickness
   of about 5\,nm was comparable with the coherence length, so vortices should not form.  This was
  confirmed by the fact that no marked difference between the spectra
  with terahertz electric field set perpendicular and parallel to the external
  magnetic field was observed. Simultaneous use of Maxwell-Garnett effective medium theory and the model of
  optical conductivity by Herman and Hlubina proved to correctly reproduce the
  terahertz spectra obtained experimentally in a magnetic field of up to 7\,T. This let us
  conclude that the magnetic field tends to suppress the superconductivity, resulting in an inhomogeneous state where
  superconducting domains are enclosed within a normal-state
  matrix. The scattering rate due to pair-breaking effects was found
  to linearly increase with magnetic field. 
  \end{abstract}



  \maketitle

  \section{Introduction}

  In superconducting materials, the interactions between external static
  magnetic field and charge carriers are of fundamental importance.
  Magnetic field induces screening currents (orbital effect), and it interacts with the electron spin (Zeeman effect).
  Screening currents usually suppress the magnetic field in the bulk, so the
  Zeeman effect is weakened. However, in-plane magnetic field will fully
  penetrate films whose thickness is much lower than the penetration depth
  $\lambda$; therefore, the Zeeman effect can be dominant.
  In case of low spin-orbit scattering, magnetic field shifts the density of
  states (DOS) of electrons with spins up and down by $\pm \mu_{\rm B} \mu_0
  H$, where $H$ is the magnetic field intensity, $\mu_0$ is the
  permeability of free space and $\mu_{\rm B}=9.274 \times 10^{-24}\,\rm J \cdot
  T^{-1}$ denotes the Bohr magneton. 
  The spin-orbit scattering leads to spin flipping, and the
  spectroscopic gap $\Omega_G$ is reduced by $2 \mu_{\rm B} \mu_0 H$. As the
  spin-orbit scattering rate increases, peaks in
  the DOS due to up and down spins are smeared and,
  eventually, only one broad peak in the DOS remains.
  The spin-orbit scattering rate is proportional to the fourth power of atomic
  number. Therefore, low spin-orbit scattering is typical of atoms with a low atomic
  number, which was confirmed by observations in Al films ~\cite{Meservey1970}.
  The field also breaks the time-reversal symmetry, causing an overall
  weakening of the superconducting state. Abrikosov and Gorkov derived a
  theory describing superconductivity in the presence of magnetic impurities
  involving a
  pair-breaking parameter $\alpha$~\cite{Abrikosov1961}. Later,
  Maki~\cite{Maki1969} and de Gennes~\cite{deGennes1964}
  proved that all ergodic pair-breaking perturbations contribute to
  the pair-breaking parameter $\alpha$. This applies also to
  ultrathin superconducting films---those having a thickness comparable with or smaller than the coherence
  length---in the dirty limit, exposed to either
  in-plane or out-of-plane magnetic fields.
   These become effectively two-dimensional, and an in-plane
  magnetic field can induce a superconductor-insulator transition
  (SIT) \cite{Gantmakher2000,Parendo2006}. The SIT introduces a nanoscale
  inhomogeneity, causing either an enhancement of Coulombic
  interactions and a gradual decrease in the superconducting gap
  energy (fermionic scenario \cite{Finkelstein87}), or a gradual loss of coherence
  within the condensate (bosonic scenario \cite{Fisher89}).
  Such inhomogeneities were reported for various cases of the SIT~\cite{Gantmakher2000,Sacepe2008,Noat2013}.

  In order to extend the knowledge of the behavior of ultrathin
  superconducting films, we studied the effects of in-plane magnetic field on
  the optical conductivity by means of time-domain
  terahertz (THz) spectroscopy. We employed a custom-made experimental setup 
  with external magnetic field. Our measurements are able to provide access to
  the features which characterize the
  superconductivity, namely density of states, quasiparticle and Cooper-pair
  concentrations, vortex dynamics~\cite{Dressel2013}, and mesoscopic inhomogeneity.
  In the present work, we focus on the thinnest available NbN film
  from the series studied in Ref.~\onlinecite{Sindler2018} which shows an onset of the SIT.

  \section{Experiment}

  The film was deposited on a $10\times10\times1\,\rm mm^3$-sized (100) MgO
  substrate by reactive magnetron sputtering of a 99.999\%-pure Nb
  target in a mixed Ar / N$_2$ atmosphere with partial pressures of $P_\mathrm{Ar} =
  1.5\times10^{-3}\,\mathrm{mbar}$ and $P_\mathrm{N_2} =
  3.3\times10^{-4}\,\mathrm{mbar}$. The substrate holder was heated to
  $850\,^{\circ}\mathrm{C}$, and NbN was deposited at a rate of
  $\sim0.12$~nm/s. More details about the deposition technology can be
    found in Ref.~\onlinecite{Henrich2012}. The thickness of our film was
      determined from the known deposition rate as $d= 5.3$\,nm which is close to the
   typical coherence length in NbN (4--7\,nm~\cite{Xithesis}) and much
   less than the penetration depth $\lambda=2.3 \times 10^{-7}$\,m, as estimated from the imaginary part of complex conductivity at low frequencies. The critical
    temperature $T_{\rm c}=13.9\,\rm K$ was obtained in an 
    independent DC-resistivity measurement; this value is slightly lower than in the
  thicker samples from the same series (15.2 K and 15.5 K for 14.5 nm and  30.1 nm
  thick samples, respectively)
     \cite{Sindler2018}. We note that the critical temperature of NbN can
     reach up to 17.3\,K~\cite{Poole}.

   THz spectroscopy experiments consisted in measuring the sample 
  transmittance using a custom-made time-domain spectrometer. Broadband THz pulses were 
  generated using a Ti:sapphire femtosecond laser (Vitesse, Coherent) and
  a 
  large-area interdigited semiconductor emitter (TeraSED, GigaOptics). The 
  sample was placed in an Oxford Instruments Spectromag He-bath cryostat 
  with mylar windows and a superconducting coil, allowing for cooling the sample down 
  to $T=2\,\mbox{K}$. The Voigt geometry---i.e., external static magnetic
  field directed along the sample plane---was used, 
  and the  magnetic field was varied up to the maximum
  value of $\mu_{0}H= 7\,\mbox{T}$. 
  The electric vector of the linearly polarized THz pulses was set either parallel
  or perpendicular to $\bm H$.\@ The transmitted time profiles of electric field
  intensity $E(t)$
  were detected by phase-sensitive electro-optic sampling \cite{Nahata1999} in a 1\,mm thick 
  $\langle110\rangle$ ZnTe crystal. The frequency~($\nu$) dependence of complex transmittance $\tilde{t}(\nu)$ was evaluated as
  the ratio between Fourier transforms $E_{\rm s}(\nu)$ and $E_{\rm r}(\nu)$ of
  the time profiles transmitted through the sample 
   and a bare MgO reference substrate, respectively; this approach is
   known to effectively eliminate all instrumental functions.
   Prior to the numerical computations of the complex conductivity of the film
   $\tilde\sigma(\nu)$, interference effects in the MgO
   substrates were avoided by
   truncating the measured time profiles $E(t)$ before the echoes arising from internal reflections,
   whereas interferences in the thin film were accounted for \cite{poznamka}:
  \begin{equation}
  \label{t_formula}
  \tilde{t}(\nu)=\frac{E_{\rm s}(\nu)}{E_{\rm r}(\nu)}=\frac{\left[ 
    1+\tilde{n}_{\rm
   sub}(\nu)\right] \mbox{e}^{\rm i \psi(\nu)}}{1+\tilde{n}_{\rm sub}(\nu)+Z_0 \tilde{\sigma}(\nu)d},
  \end{equation}
  where $Z_0$
  denotes the vacuum impedance,  $\tilde{n}_{\rm sub}(\nu)$ the complex refractive index of the
  substrate, and $\psi(\nu)$ is the
  phase delay due to different optical thicknesses of the sample and
  reference. Since, in both these cases, the
  same substrate material (MgO) was used, we can write
  \begin{equation}
    \psi(\nu) = \frac{2\pi\nu}{c}\left[(\tilde{n}(\nu) - 1)d+\tilde{n}_{\rm sub}(\nu)(d_{\rm sub}-d_{\rm sub}^{\rm r})\right],
  \end{equation}
   where $c$ is the light velocity, $\tilde{n}$ is the
   complex refractive index of the film, $d_{\rm sub}$ and $d_{\rm sub}^{\rm r}$ are
   the thicknesses of the sample-supporting and
   reference substrates, respectively. The first term in the square brackets stems from  the propagation in
   the film, and the second one reflects the
   different thicknesses of the sample and reference
   substrates.

  Alternatively, the complex conductivity of the film in the superconducting
  state $\tilde\sigma^{\rm sc}(\nu)$ can be obtained by
   assuming a precise knowledge of the film
  conductivity $\tilde\sigma^{\rm n}(\nu)$ in the normal state
  just above $T_{\rm c}$.
  In fact, as one evaluates the ratio of transmittances in
    the superconducting and normal states using Eq.~(\ref{t_formula}),
  the numerators cancel out, yielding:
  \begin{equation}
  \label{t_formula2}
  \frac{\tilde{t}_{\rm sc}(\nu)}{\tilde{t}_{\rm n}(\nu)}=\frac{1+\tilde{n}_{\rm
    sub}(\nu)+Z_0
  \tilde{\sigma}^{\rm n}(\nu) d}{1+\tilde{n}_{\rm
  sub}(\nu)+Z_0 \tilde{\sigma}^{\rm sc}(\nu)d}
  \end{equation}
  from which $\tilde\sigma^{\rm sc}(\nu)$ can be evaluated. This
  simplification is  possible because the optical properties of the MgO
  substrate change very weakly within such
  a narrow temperature interval. As an advantage, this approach 
  is not affected by any inaccuracy in the values of $d_{\rm sub}$, $d_{\rm
  sub}^{\rm r}$.  The normal-state
  conductivity $\tilde\sigma^{\rm n}(\nu)$ was
  carefully determined previously \cite{Sindler2018}.

  \section{Results and discussion}

  In the present work, the sample was probed by linearly polarized THz pulses in two distinct
  geometries, with
  electric field vector $\bm E$ parallel and perpendicular to the direction of in-plane magnetic
  field, which we mark in the following by symbols $E^{\parallel}$ and
  $E^{\perp}$, respectively. The complex conductivity of the film $\tilde{\sigma}(\nu)$ was evaluated by the two above-described methods 
  based on Eqs.~(\ref{t_formula}) and (\ref{t_formula2}).

  \subsection{Zero magnetic field} We start with the analysis of normal-state properties of our ultrathin NbN film.
  They were found to be accurately described by the Drude model,
  yielding a DC conductivity of $\sigma_0=(1.53 \pm 0.02) \,\mathnormal{\upmu}\Omega^{-1} \rm m^{-1}$  and a
  scattering time  $\tau_{\rm n}=(15\pm 8)$\,fs \cite{Sindler2018}. The latter value
  indicates a moderate disorder, and, in
  principle, quantum corrections to the
  Drude model might be applicable \cite{Altshuler1987,Neilinger2019}.
  However,  Cheng
  \textit{et al.} \cite{Cheng2016} systematically studied
  NbN films with an increasing level of disorder, and quantum corrections
  appeared to play no significant role in their best-quality
  samples, exhibiting the highest values of $T_{\rm c}$.
  Since the value of $T_{\rm c}$ in our film even slightly exceeds those reported in
  Ref.~\onlinecite{Cheng2016}, we conclude that quantum corrections need not be
  taken into account.

  In general, it is quite difficult to determine precisely the THz
  properties of superconducting films. This is especially true of the
  transmittance phase, because one has to distinguish between the
  contributions of the thin film and that of the
  substrate which is much thicker, in our case by more than five
  orders of magnitude thicker.
  We optimized the substrate thickness within the experimental
  accuracy so that THz spectra for $H=0$ show no dissipation [$\sigma_1(\nu)\approx
  0$] below the 
  optical gap $2\Delta$ in the zero-temperature limit. This corresponds to the expected
  behavior. By numerical
  calculations using Eq.~(\ref{t_formula}), the precise mean value
  of the substrate thickness  was found as $d_{\rm sub}=(990.0\pm 0.5) \,\upmu\rm
  m$. The alternative method using Eq.~(\ref{t_formula2})
   lead to the same results, see  Fig.~\ref{fig:compare} ($\sigma_2(\nu)$ is shown
   in the Supplemental Material \cite{SupMat}).

  \begin{figure}[t]
  \centering
  \includegraphics[angle=270,width=\columnwidth]{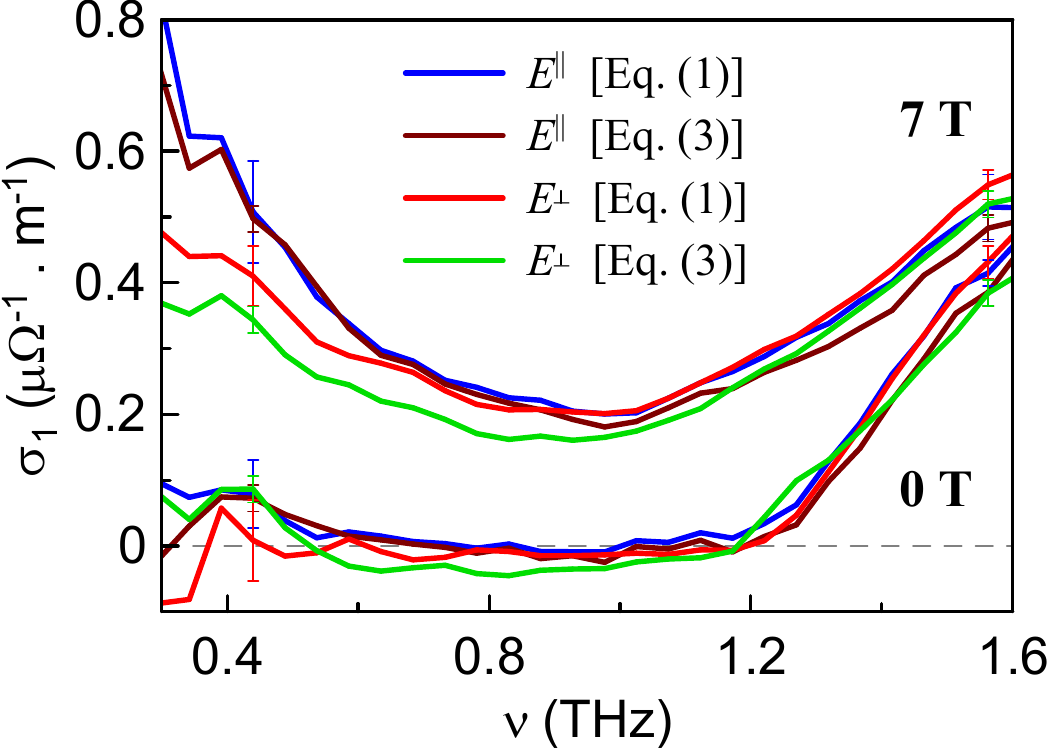}
  \caption{
    Real part spectra of NbN conductivity $\sigma_1(\nu)$ obtained by
    the two described methods from experiments using both linear
  polarizations at $T=3\,\rm K$ and a magnetic field of 0 and 7\,T. Note that the evaluation of the
  imaginary part $\sigma_2(\nu)$ provides spectra with a much lower uncertainty.
  }
  \label{fig:compare}
  \end{figure}

    At zero magnetic field, the sample conductivity shows features typical
    of a classical BCS superconductor---below the optical gap $2\Delta(0)$,
    there is no dissipation in the real part,
    i.e., $\sigma_1\approx 0$, and a $1/\nu$ dependence in the imaginary part
    $\sigma_2(\nu)$ is observed (see black lines in Fig.~\ref{fig:Bexp}).
  At low frequencies, $\sigma_1 \ll \sigma_2$; thus $\sigma_1$ is determined less
  reliably than $\sigma_2$.

  The zero-field conductivity can be best described by the
  Herman-Hlubina (HH) model for Dynes
  superconductors~\cite{Herman2017}, see the red lines in
  Fig.~\ref{fig:Bfit}.
  Within this model, the scattering rate in the normal state $\Gamma_{\rm n}$
  can be written as a sum of two terms related to the superconducting state:
  $\Gamma_{\rm n}=\Gamma_{\rm s}+\Gamma$ where $\Gamma_{\rm s}$ and $\Gamma$ are the scattering
    rates relative to the pair-conserving and pair-breaking
    processes, respectively. The parameter $\Gamma$ also appears in
    the well-known Dynes formula~\cite{Dynes1978} where it accounts for
    the broadening of the DOS which  diverges at
    $\pm \Delta$ for $\Gamma=0$.
  The normal-state scattering rate is linked to the scattering
    time $\tau_{\rm n}$ by the relation $\Gamma_{\rm n}= (h/2\pi) (2\tau_{\rm n})^{-1}$, where $h$
    is the Planck constant.
  In the limit $\Gamma=0$, i.e., without pair-breaking processes, the HH model
  reproduces the Zimmermann model \cite{Zimmermann1991} which was
  successfully applied in the previous report on the ultrathin NbN sample
  \cite{Sindler2018}. 
  In the present work, by fitting our zero-field spectra of the complex
  conductivity $\tilde{\sigma}(\nu)$ using
  the HH model, we found $\Gamma_{\rm s}/h = (5.6\pm 0.4)$\,THz~\cite{THz}, $\Gamma/h \leq
  10^{-3}$\,THz and $2\Delta(0)/h=1.2$\,THz. Due to the small value of $\Gamma$,
  the Zimmermann model is fully appropriate for the magnetic-field-free
  case. 
  Indeed, from the values of scattering rates found in the superconducting state,  
  we obtained $\tau_n=(14\pm1)$\,fs which is in excellent agreement with the
  Drude fit of the normal-state spectra reported in
    Ref.~\onlinecite{Sindler2018}. However, unlike the HH model, the Zimmermann model is not able to account for
  modifications of the superconducting state due to magnetic
  field.

  \subsection{Magnetic field dependence}

\begin{figure}[t]
  \centering
  \includegraphics[width=\columnwidth]{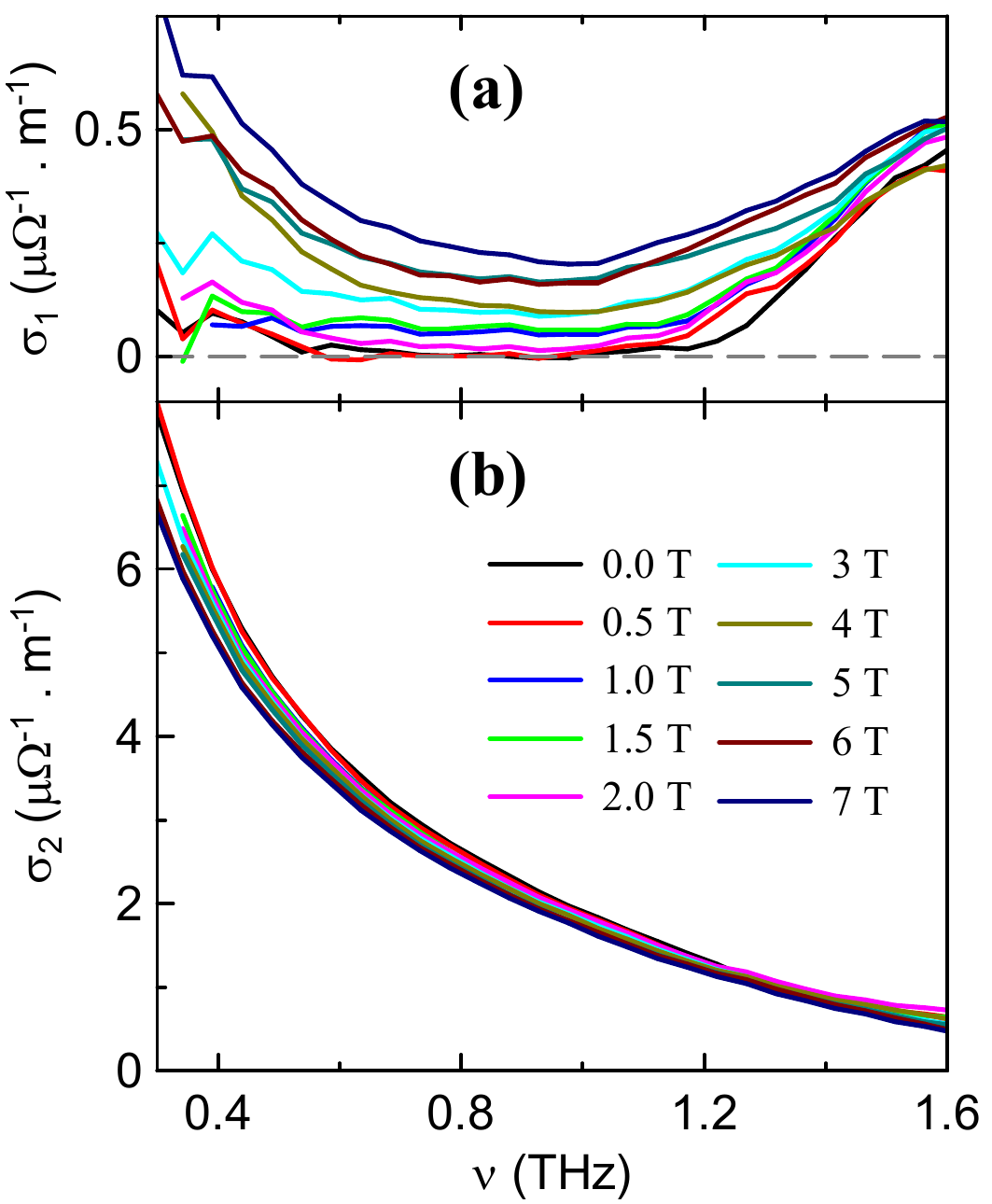}

  \caption{
  Real (a) and imaginary (b) part of the conductivity of the NbN film
  under varying in-plane magnetic field at $T= 3\,K$ for E$^\parallel$
  evaluated using Eq.~(\ref{t_formula}).
  }
  \label{fig:Bexp}
  \end{figure}

  Upon applying the in-plane magnetic field, the complex conductivity is modified, see Fig.~\ref{fig:Bexp}.
  Whereas the imaginary-part spectra $\sigma_2(\nu)$ vary only
  slightly even for the highest attainable field, the
  real-part spectra $\sigma_1(\nu)$ show more significant changes.
  The dissipation increases with magnetic field; this occurs
  especially at low frequencies where an upward tail in $\sigma_1(\nu)$ gradually
  develops. We did not observe any significant differences between the
  spectra corresponding to the THz pulses with linear polarizations
  parallel and perpendicular to the external magnetic field.
  By contrast,
  similar experiments with a thicker NbN sample revealed a
  strongly polarization-dependent transmittance~\cite{Sindler2017}.
  In the present case, the
  fact that the spectra for the two THz polarization directions are
similar lets us
  conclude that the NbN sample contains no vortices oriented along the
  direction of the magnetic field, in contrast to predictions for thicker
  samples under in-plane applied magnetic field \cite{Luzhbin2001}. This is
  in agreement with our estimate that the coherence length is
  comparable with the film thickness, thus the vortex cores
  do not fit in.

  Our experimental observations are different from those of
  Xi \textit{et al.}~\cite{Xi2010,Xithesis} who reported a prominent
  decrease in the gap energy upon 
  increasing magnetic field, whereas there was no dissipation below
  the gap.  They employed successfully the model of complex conductivity developed by
  Skalski \textit{et al.}~\cite{Skalski1964}. Thus, we conclude that this model is not applicable in our case. The
  NbN sample studied by Xi \textit{et al.} was 70\,nm thick, it had a critical
  temperature of $T_{\rm c}=12.8$\,K, and a scattering time
  of $\tau_{\rm n}=0.07$\,fs, two orders of magnitude lower than in the
  present case. We suppose that the major difference in the
  spectral responses is linked to the sample thicknesses; whereas the sample
  in Ref.~\cite{Xi2010} was much thicker than the typical coherence
  length,
  the thickness of our sample was comparable with the coherence length.
  Consequently, only our sample can be considered as a two-dimensional system.

  We assume that upon
  applying in-plane magnetic field, the superconducting properties are
  weakened by pair-breaking processes and, additionally, the system becomes inhomogeneous.  In
  such a case, within the
  film, isolated superconducting islands are formed, surrounded by a
  matrix in which the superconducting properties are heavily
  suppressed; this part can be approximated by the 
  normal state. The spectral response of superconducting islands is
  described by the HH model~\cite{Herman2017} where $\Gamma$
  plays a role analogous to the pair-breaking parameter $\alpha$. We
  have reproduced the experimental data by the
  Maxwell-Garnett model~\cite{Garnett1904} in which the superconducting islands
  were treated as particles and the normal state as a matrix~\cite{Sindler2018}:

  \begin{equation}\label{mgt}
   \frac{\tilde{\sigma}_\mathrm{MG}(\nu)  - \tilde{\sigma}_\mathrm{n}(\nu) }{L
   \tilde{\sigma}_\mathrm{MG}(\nu) + (1-L)  \tilde{\sigma}_\mathrm{n}(\nu)} =
   f_\mathrm{s} \frac{\tilde{\sigma}_\mathrm{s}(\nu)-\tilde{\sigma}_\mathrm{n}(\nu)} {L\tilde{\sigma}_\mathrm{s}(\nu)+(1-L) \tilde{\sigma}_\mathrm{n}(\nu)},
  \end{equation}
  where $f_{\rm s}$ and $L$ are the volume fraction
    and the depolarization
    factor of the superconducting inclusions, respectively. Although the
    actual topology of the superconducting film can be non-trivial, we assume
    the depolarization factor to amount to $L=1/3$. This value is usually
    employed for calculating the response of flat disks embedded in a matrix,
  but it may equally describe other geometries.

  The use of the Maxwell-Garnett formula for high concentrations of
  inclusions  might be questioned. However, Rychetsk\'y \textit{et al.}\@
  \cite{Rychetsky2004} argued that the Maxwell-Garnett formula holds even
  for high concentrations of inclusions as long as the matrix is percolated.
  In our fits, we used two fitting parameters---the volume
  fraction of superconducting islands $f_{\rm s}$, and the
  pair-breaking rate $\Gamma$ which determines the shape of the $\sigma_s(\nu)$ spectra.
  In fact, varying $\Gamma_s$ and $\Delta$ did not improve our fits, so we conclude that
  their values are independent of the magnetic field.

  \begin{figure}[t]
  \centering
  \includegraphics[width=0.45\textwidth]{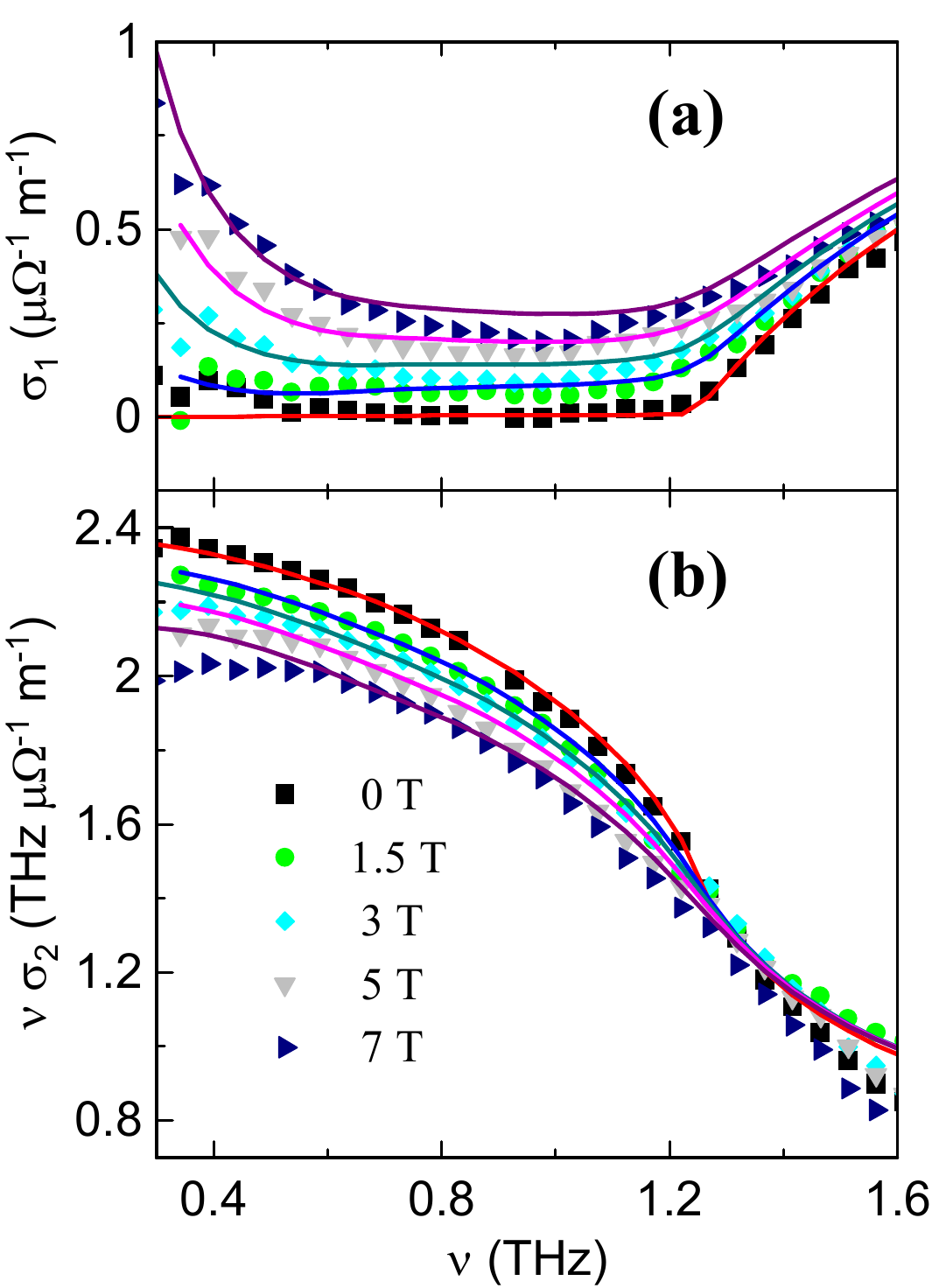}
  \caption{
    Symbols: real (a) and imaginary (b) part of the conductivity of NbN at $T=3\,\rm K$
    for $E^{\parallel}$ evaluated using Eq.~(\ref{t_formula}). Lines are fits by the Maxwell-Garnett formula [Eq.~(\ref{mgt})] using 
    the Drude model for the normal-state
    $\tilde{\sigma}_{\rm n}(\nu)$ and the HH model for the
    superconducting-state $\tilde{\sigma}_{\rm s}(\nu)$ components,
    respectively. The imaginary part is plotted as $\nu
  \sigma_2(\nu)$ for improved clarity. 
  }
  \label{fig:Bfit}
  \end{figure}

  \begin{figure}[h]
  \centering
  \includegraphics[width=0.45\textwidth]{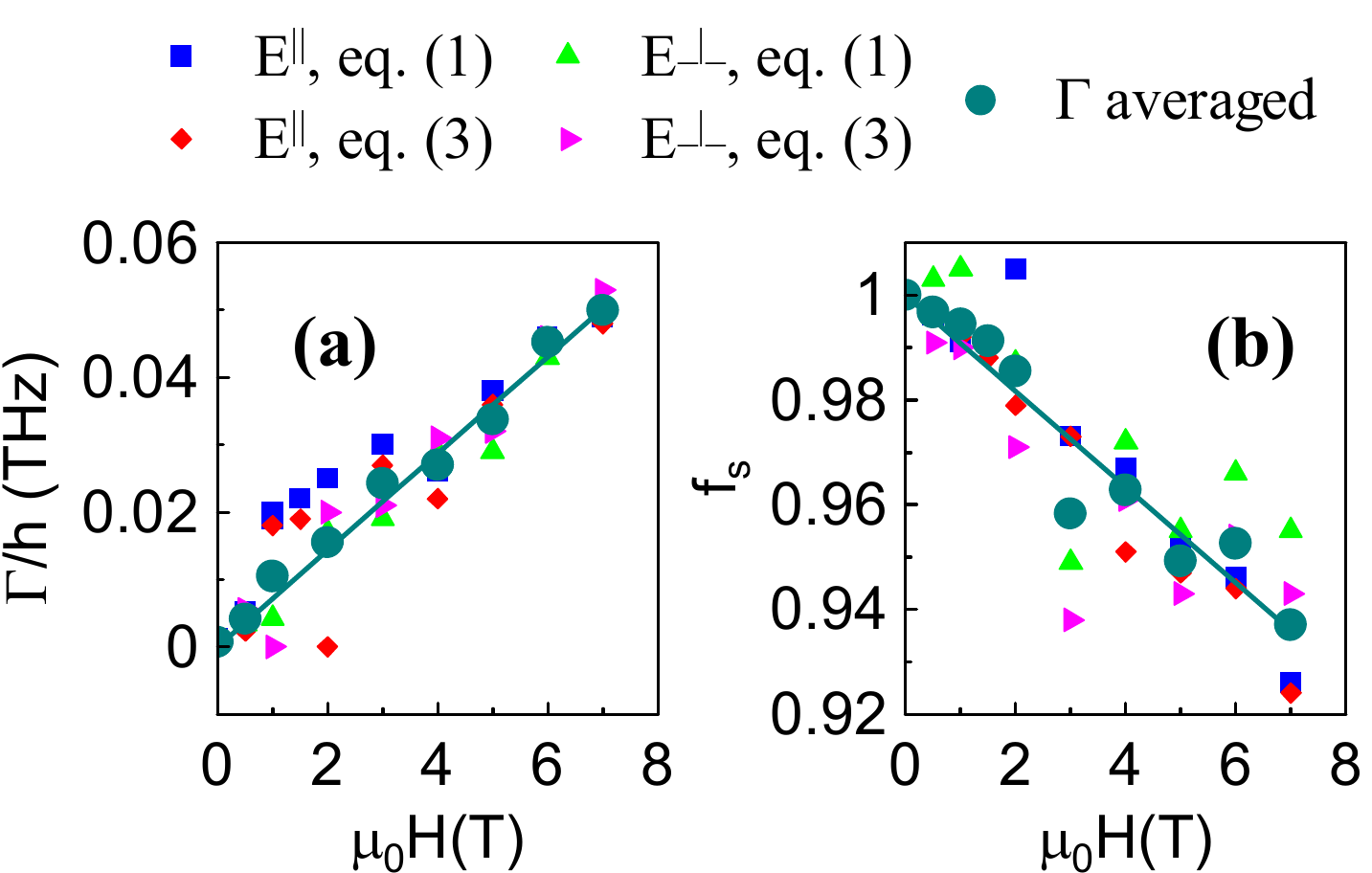}

  \caption{Symbols: (a) Cooper-pair-breaking scattering rate $\Gamma$ and (b) volume
  fraction of superconducting inclusions $f_{\rm s}$ obtained from
  fits of $\tilde{\sigma}(\nu)$ spectra at $T=3\,\rm K$, as a function of applied in-plane
  magnetic field. $\Gamma$ and $f_{\rm s}$ are the only
  two free parameters of the used Maxwell-Garnett model in combination with
  the HH and Drude models. Lines: fits demonstrating
  that both these parameters depend linearly on the magnetic field.
  }
  \label{fig:parameters}
  \end{figure}

  Our model describes correctly all observed spectral features, see
  Fig.~\ref{fig:Bfit}. THz conductivity spectra for both linear
  polarizations \textit{evaluated by both methods} were fitted using only two free
  parameters: $\Gamma$ and $f_{\rm s}$, see Fig.~\ref{fig:parameters}.  Although
  there is some scatter in the parameters, their field dependences exhibit
  clear trends. Whereas  the volume fraction of superconducting inclusions
  $f_{\rm s}$ linearly decreases with $H$ from 1 to 0.94,
  the pair-breaking scattering rate $\Gamma$ linearly rises
  with the magnetic field as
  $\Gamma/h=0.0072\,\mbox{THz}\cdot\mbox{T}^{-1} \mu_0 H$, which corresponds to
  $0.5 \mu_B \mu_o H/h$. The field-dependences of $f_s$ and
  $\Gamma$ would be qualitatively the same if we assumed different values of
  $L$; however, the experiment provides no means to obtain its most
  appropriate value.

  The field-dependences $\Gamma(H)$ of the pair-breaking scattering
  rate upon a strong spin-orbit
  interaction were predicted to be quadratic for an in-plane magnetic
  field and linear for an out-of-plane field~\cite{Tinkham1996,Fulde2010}. 
  Xi \emph{et al.}~\cite{Xi2013} performed far-infrared measurements under
  an in-plane magnetic field, and they evaluated the
pair-breaking parameter using the Skalski model~\cite{Skalski1964}.
They obtained, in agreement with the theory, a quadratic dependence for
a NbTiN thin film, but they also
observed a linear behavior for their NbN film with
$\Gamma/h=0.014\,\mbox{THz}\cdot\mbox{T}^{-1}\mu_O/h= \mu_B \mu_0 H/h$. In contrast
with our results, they observed no absorption below the
optical gap, and they found a linear decrease in the
spectroscopic gap $\Omega_G$ with quite a steep slope of
$-0.12\,\mbox{THz}\cdot\mbox{T}^{-1}$ ($\sim 8.6 \mu_B/h$).  This is
different from the trend observed in our data which revealed only
a small decrease in the optical gap even at the magnetic field of 7\,T. 

  We believe that the observed linear dependence $\Gamma(H)$
  is due to a combination of Zeeman splitting of the DOS and the
  spin-orbit interaction. The peaks in the DOS become doublets due to the Zeeman
  effect, and they merge into a broad peak with a shape which can be well described by
  a formula pertinent for a Dynes superconductor. 

  \subsection{Temperature dependence under magnetic field}
\begin{figure}
  \centering
  \includegraphics[width=0.45\textwidth]{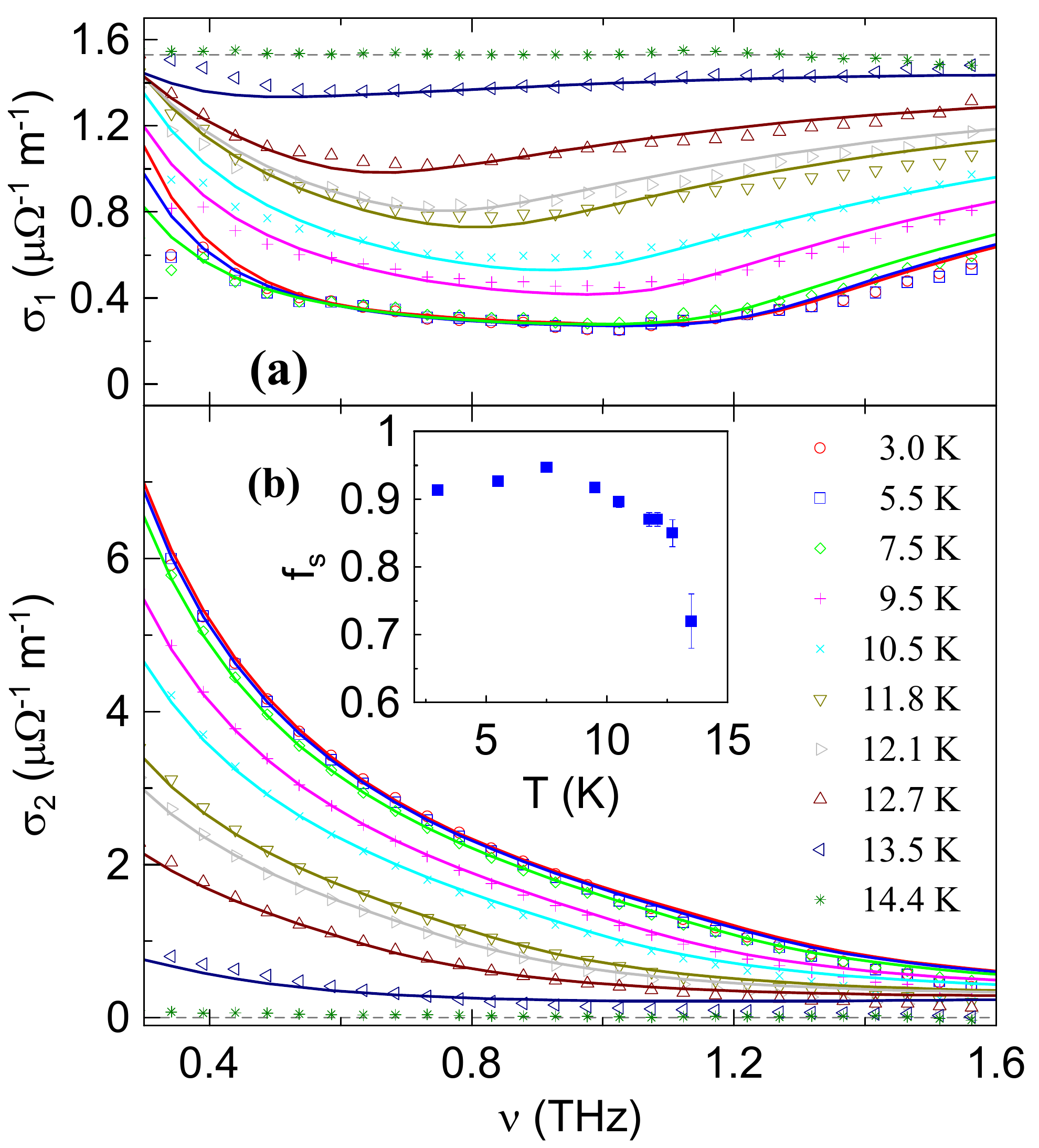}

  \caption{Symbols: real (a) and imaginary (b) conductivity of NbN at $\mu_0H=7\,\rm T$
   for linear polarization parallel with the external magnetic field evaluated from the transmission ratio
  $t_{\rm sc}/t_{\rm n}$. Lines: fits using the
  Maxwell-Garnett [Eq.~(\ref{mgt})], HH (Ref.~\onlinecite{Herman2017}), and Drude
  models. In the fitting, we used
  a temperature-dependent gap energy $\Delta(T)$ expressed by
  Eq.~(\ref{gap_t}), a temperature-independent scattering rate $\Gamma$,
  whereas the superconducting fraction $f_{\rm s} (T)$ was used as a
  free fitting parameter at each temperature. Inset: temperature dependence
  of the superconducting fraction $f_{\rm s}$ obtained from the fits.
  }
  \label{fig:sigma7T}
  \end{figure}

  In order to further test the
  validity of our model, we
  performed an additional set of measurements. It consisted in setting the
  magnetic field to the highest attainable value of $\mu_0H=7\,\rm T$ and
  measuring the temperature
  dependence of the conductivity of the NbN film. The experimental results
  (shown by symbols in Fig.~\ref{fig:sigma7T}) were evaluated from the transmission ratio $\tilde{t}_{\rm
  sc}/\tilde{t}_{\rm n}$ (see Eq.~\ref{t_formula2}) where the normal-state transmittance was measured at
  $T=14.4\,\rm K$. In the fitting procedure,
   in order to account for the effect of heating, we assumed the
   temperature dependence of the gap in the form \cite{Sheahen1966}
  \begin{equation}
  \label{gap_t}
  \Delta(T)=\Delta(0) \sqrt{\cos \frac{\pi}{2} \left( \frac{T}{T_{\rm c}}\right)^2}.
  \end{equation}
  The scattering rate $\Gamma$ was assumed to be temperature independent
  in agreement with previous measurements of different NbN sample~\cite{Sindler2014}. 
  In a MoC film, which is a similar superconducting compound, $\Gamma$ was found to
  be temperature independent up to 0.5 $T_c$ 
  and at higher temperatures, it reached up to four times the low-temperature value~\cite{Szabo2016}.
  Our model shows an excellent agreement with the experiment, see Fig.~\ref{fig:sigma7T}.
  This lets us conclude that at fixed magnetic field of $\mu_0H=7\,\rm T$,
  the superconducting fraction $f_{\rm s}$ is only weakly temperature
  dependent up to $T=10$\,K,
  and it sharply decreases upon further heating (see inset of Fig.~\ref{fig:sigma7T}).

  \section{Summary and conclusion}
  We studied the optical conductivity of a high-quality ultrathin NbN
  film under in-plane applied external
  magnetic field of up to 7\,T, in the low-temperature limit. We measured its
  transmission in the THz range (0.4--1.6\,THz), utilizing broadband
  pulses with linear polarization set either parallel
  or perpendicular to the applied static magnetic field, and we
  evaluated the complex conductivity spectra $\tilde{\sigma}(\nu)$ using two 
  different methods, one  based on a direct computation [Eq.~(\ref{t_formula})]  and one employing the ratio between the complex transmittances of the superconducting and normal states $\tilde{t}_{\rm sc}(\nu)/\tilde{t}_{\rm n}(\nu)$ [Eq.~(\ref{t_formula2})].
  Both these approaches lead essentially to the same results,
  as shown in Fig.~\ref{fig:compare}. Nevertheless, $\tilde{\sigma}(\nu)$ can be determined more precisely by the latter method, since the main source of experimental uncertainty originates in the
  substrate thickness $d_{\rm sub}$ which is effectively canceled in the ratio, and the
  relative errors in determining conductivity values are lower in the normal state than
  in the superconducting one.

  The zero-magnetic field spectrum is well described by the Zimmermann
  model~\cite{Zimmermann1991},
  as discussed in our previous publication~\cite{Sindler2018}. In the present
  case, however,  we used a more general model---that for Dynes superconductors
  proposed by Herman and
   Hlubina~\cite{Herman2017}, which is able to take into account the
   Cooper-pair-breaking processes.
  We found that the scattering rates in the normal and the superconducting
  states are in excellent agreement
     ($\Gamma_{\rm s}=\Gamma_{\rm n}$). The
     value of $\Gamma$ is negligibly small, suggesting that Cooper-pair-breaking
     processes are weak, thus confirming the applicability of the Zimmermann model.

     In the low-temperature limit, we observed a significant modification of the conductivity under in-plane magnetic field. 
   Its imaginary part is dominated by the London term, $\sigma_2(\nu)
   \sim 1/\nu$, and it only slightly decreases with $H$.
  The real part $\sigma_1(\nu)$ exhibits an marked absorption even at
  frequencies below the optical gap; it does not vanish even at the highest
  attained magnetic field.
  We did not observe any relevant difference between the spectra
  obtained in $E^{\parallel}$ and $E^{\perp}$ configurations, unlike 
  in our previous experiments with thicker NbN films~\cite{Sindler2017}. This
  absence of anisotropy rules out the presence of vortex chains predicted
  earlier by  Luzhbin~\cite{Luzhbin2001}.

   We found that our experimental results
  can be explained by assuming a local suppression of superconducting
   properties, resulting in an inhomogeneous state of superconducting islands in a
   normal state matrix with a complex topology.  The properties of superconducting islands are 
  modified by pair-breaking effects proportional to the strength of the
  applied magnetic field. We found that in the present case 
  the HH model for Dynes superconductors~\cite{Herman2017} is more suitable
  than the Skalski model~\cite{Skalski1964}. 
   .
   An inhomogeneity on the nanoscale
  arising owing to the SIT was reported earlier in other
  cases~\cite{Gantmakher2000,Sacepe2008,Noat2013}. In order to quantitatively describe our
    spectra, we developed 
  a model assuming superconducting islands enclosed within a
  normal-state matrix, based on the Maxwell-Garnett theory~\cite{Garnett1904}.
  The complex topology was described by assuming a depolarization
  factor of $L=\frac{1}{3}$. Our model yields a linear
  decrease in the volume fraction of superconducting islands
  $f_s$ with magnetic field.
  At the same time, we observed a gradual decrease in the
  superconducting properties of our NbN film, which is
  reflected by the linear rise in the pair-breaking scattering rate
  $\Gamma$ with magnetic field. Finally, our approach has proved to remain
  valid also for higher temperatures.

\section{Acknowledgment}
We are grateful to K.~Ilin and M.~Siegel for preparing and characterizing the NbN sample
as well as to F.~Herman and R.~Hlubina for fruitful discussions. We also acknowledge the
financial support by the Czech Science Foundation (Project No.~21-11089S) and by
the MŠMT Project No.~SOLID21 -- CZ.02.1.01/0.0/0.0/16\_019/0000760.
\bibliographystyle{aipnum4-2}
\bibliography{NbN3.4.bib}

\begin{thebibliography}{37}%
\makeatletter
\providecommand \@ifxundefined [1]{%
 \@ifx{#1\undefined}
}%
\providecommand \@ifnum [1]{%
 \ifnum #1\expandafter \@firstoftwo
 \else \expandafter \@secondoftwo
 \fi
}%
\providecommand \@ifx [1]{%
 \ifx #1\expandafter \@firstoftwo
 \else \expandafter \@secondoftwo
 \fi
}%
\providecommand \natexlab [1]{#1}%
\providecommand \enquote  [1]{``#1''}%
\providecommand \bibnamefont  [1]{#1}%
\providecommand \bibfnamefont [1]{#1}%
\providecommand \citenamefont [1]{#1}%
\providecommand \href@noop [0]{\@secondoftwo}%
\providecommand \href [0]{\begingroup \@sanitize@url \@href}%
\providecommand \@href[1]{\@@startlink{#1}\@@href}%
\providecommand \@@href[1]{\endgroup#1\@@endlink}%
\providecommand \@sanitize@url [0]{\catcode `\\12\catcode `\$12\catcode
  `\&12\catcode `\#12\catcode `\^12\catcode `\_12\catcode `\%12\relax}%
\providecommand \@@startlink[1]{}%
\providecommand \@@endlink[0]{}%
\providecommand \url  [0]{\begingroup\@sanitize@url \@url }%
\providecommand \@url [1]{\endgroup\@href {#1}{\urlprefix }}%
\providecommand \urlprefix  [0]{URL }%
\providecommand \Eprint [0]{\href }%
\providecommand \doibase [0]{https://doi.org/}%
\providecommand \selectlanguage [0]{\@gobble}%
\providecommand \bibinfo  [0]{\@secondoftwo}%
\providecommand \bibfield  [0]{\@secondoftwo}%
\providecommand \translation [1]{[#1]}%
\providecommand \BibitemOpen [0]{}%
\providecommand \bibitemStop [0]{}%
\providecommand \bibitemNoStop [0]{.\EOS\space}%
\providecommand \EOS [0]{\spacefactor3000\relax}%
\providecommand \BibitemShut  [1]{\csname bibitem#1\endcsname}%
\let\auto@bib@innerbib\@empty
\bibitem [{\citenamefont {Meservey}, \citenamefont {Tedrow},\ and\
  \citenamefont {Fulde}(1970)}]{Meservey1970}%
  \BibitemOpen
  \bibfield  {author} {\bibinfo {author} {\bibfnamefont {R.}~\bibnamefont
  {Meservey}}, \bibinfo {author} {\bibfnamefont {P.~M.}\ \bibnamefont
  {Tedrow}},\ and\ \bibinfo {author} {\bibfnamefont {P.}~\bibnamefont
  {Fulde}},\ }\href {https://doi.org/10.1103/PhysRevLett.25.1270} {\bibfield
  {journal} {\bibinfo  {journal} {Phys. Rev. Lett.}\ }\textbf {\bibinfo
  {volume} {25}},\ \bibinfo {pages} {1270} (\bibinfo {year}
  {1970})}\BibitemShut {NoStop}%
\bibitem [{\citenamefont {Abrikosov}\ and\ \citenamefont
  {Gor'kov}(1960)}]{Abrikosov1961}%
  \BibitemOpen
  \bibfield  {author} {\bibinfo {author} {\bibfnamefont {A.}~\bibnamefont
  {Abrikosov}}\ and\ \bibinfo {author} {\bibfnamefont {L.}~\bibnamefont
  {Gor'kov}},\ }\href@noop {} {\bibfield  {journal} {\bibinfo  {journal} {Zh.
  Eksp. Teor. Fiz.}\ }\textbf {\bibinfo {volume} {39}},\ \bibinfo {pages}
  {1781} (\bibinfo {year} {1960})}\BibitemShut {NoStop}%
\bibitem [{\citenamefont {Parks}(1969)}]{Maki1969}%
  \BibitemOpen
  \bibfield  {author} {\bibinfo {author} {\bibfnamefont {R.}~\bibnamefont
  {Parks}},\ }\href {https://books.google.cz/books?id=D4hK-yGBEmoC} {\emph
  {\bibinfo {title} {Superconductivity: Part 2 (In Two Parts)}}},\
  Superconductivity\ (\bibinfo  {publisher} {Taylor \& Francis},\ \bibinfo
  {year} {1969})\BibitemShut {NoStop}%
\bibitem [{\citenamefont {de~Gennes}(1964)}]{deGennes1964}%
  \BibitemOpen
  \bibfield  {author} {\bibinfo {author} {\bibfnamefont {P.~G.}\ \bibnamefont
  {de~Gennes}},\ }\href@noop {} {\bibfield  {journal} {\bibinfo  {journal}
  {Physik der kondensierten Materie}\ }\textbf {\bibinfo {volume} {3}},\
  \bibinfo {pages} {79} (\bibinfo {year} {1964})}\BibitemShut {NoStop}%
\bibitem [{\citenamefont {Gantmakher}\ \emph {et~al.}(2000)\citenamefont
  {Gantmakher}, \citenamefont {Golubkov}, \citenamefont {Dolgopolov},
  \citenamefont {Shashkin},\ and\ \citenamefont
  {Tsydynzhapov}}]{Gantmakher2000}%
  \BibitemOpen
  \bibfield  {author} {\bibinfo {author} {\bibfnamefont {V.~F.}\ \bibnamefont
  {Gantmakher}}, \bibinfo {author} {\bibfnamefont {M.~V.}\ \bibnamefont
  {Golubkov}}, \bibinfo {author} {\bibfnamefont {V.~T.}\ \bibnamefont
  {Dolgopolov}}, \bibinfo {author} {\bibfnamefont {A.}~\bibnamefont
  {Shashkin}},\ and\ \bibinfo {author} {\bibfnamefont {G.~E.}\ \bibnamefont
  {Tsydynzhapov}},\ }\href@noop {} {\bibfield  {journal} {\bibinfo  {journal}
  {J. Exp. Theor. Phys. Lett.}\ }\textbf {\bibinfo {volume} {71}},\ \bibinfo
  {pages} {473} (\bibinfo {year} {2000})}\BibitemShut {NoStop}%
\bibitem [{\citenamefont {Parendo}, \citenamefont {Tan},\ and\ \citenamefont
  {Goldman}(2006)}]{Parendo2006}%
  \BibitemOpen
  \bibfield  {author} {\bibinfo {author} {\bibfnamefont {K.~A.}\ \bibnamefont
  {Parendo}}, \bibinfo {author} {\bibfnamefont {K.~H. S.~B.}\ \bibnamefont
  {Tan}},\ and\ \bibinfo {author} {\bibfnamefont {A.~M.}\ \bibnamefont
  {Goldman}},\ }\href {https://doi.org/10.1103/PhysRevB.73.174527} {\bibfield
  {journal} {\bibinfo  {journal} {Phys. Rev. B}\ }\textbf {\bibinfo {volume}
  {73}},\ \bibinfo {pages} {174527} (\bibinfo {year} {2006})}\BibitemShut
  {NoStop}%
\bibitem [{\citenamefont {Finkelstein}(1987)}]{Finkelstein87}%
  \BibitemOpen
  \bibfield  {author} {\bibinfo {author} {\bibfnamefont {A.}~\bibnamefont
  {Finkelstein}},\ }\href@noop {} {\bibfield  {journal} {\bibinfo  {journal}
  {JETP Lett.}\ }\textbf {\bibinfo {volume} {45}},\ \bibinfo {pages} {46}
  (\bibinfo {year} {1987})}\BibitemShut {NoStop}%
\bibitem [{\citenamefont {Fisher}\ \emph {et~al.}(1989)\citenamefont {Fisher},
  \citenamefont {Weichman}, \citenamefont {Grinstein},\ and\ \citenamefont
  {Fisher}}]{Fisher89}%
  \BibitemOpen
  \bibfield  {author} {\bibinfo {author} {\bibfnamefont {M.~P.~A.}\
  \bibnamefont {Fisher}}, \bibinfo {author} {\bibfnamefont {P.~B.}\
  \bibnamefont {Weichman}}, \bibinfo {author} {\bibfnamefont {G.}~\bibnamefont
  {Grinstein}},\ and\ \bibinfo {author} {\bibfnamefont {D.~S.}\ \bibnamefont
  {Fisher}},\ }\href {https://doi.org/10.1103/PhysRevB.40.546} {\bibfield
  {journal} {\bibinfo  {journal} {Phys. Rev. B}\ }\textbf {\bibinfo {volume}
  {40}},\ \bibinfo {pages} {546} (\bibinfo {year} {1989})}\BibitemShut
  {NoStop}%
\bibitem [{\citenamefont {Sac\'ep\'e}\ \emph {et~al.}(2008)\citenamefont
  {Sac\'ep\'e}, \citenamefont {Chapelier}, \citenamefont {Baturina},
  \citenamefont {Vinokur}, \citenamefont {Baklanov},\ and\ \citenamefont
  {Sanquer}}]{Sacepe2008}%
  \BibitemOpen
  \bibfield  {author} {\bibinfo {author} {\bibfnamefont {B.}~\bibnamefont
  {Sac\'ep\'e}}, \bibinfo {author} {\bibfnamefont {C.}~\bibnamefont
  {Chapelier}}, \bibinfo {author} {\bibfnamefont {T.~I.}\ \bibnamefont
  {Baturina}}, \bibinfo {author} {\bibfnamefont {V.~M.}\ \bibnamefont
  {Vinokur}}, \bibinfo {author} {\bibfnamefont {M.~R.}\ \bibnamefont
  {Baklanov}},\ and\ \bibinfo {author} {\bibfnamefont {M.}~\bibnamefont
  {Sanquer}},\ }\href@noop {} {\bibfield  {journal} {\bibinfo  {journal} {Phys.
  Rev. Lett.}\ }\textbf {\bibinfo {volume} {101}},\ \bibinfo {pages} {157006}
  (\bibinfo {year} {2008})}\BibitemShut {NoStop}%
\bibitem [{\citenamefont {Noat}\ \emph {et~al.}(2013)\citenamefont {Noat},
  \citenamefont {Cherkez}, \citenamefont {Brun}, \citenamefont {Cren},
  \citenamefont {Carbillet}, \citenamefont {Debontridder}, \citenamefont
  {Ilin}, \citenamefont {Siegel}, \citenamefont {Semenov}, \citenamefont
  {H\"ubers},\ and\ \citenamefont {Roditchev}}]{Noat2013}%
  \BibitemOpen
  \bibfield  {author} {\bibinfo {author} {\bibfnamefont {Y.}~\bibnamefont
  {Noat}}, \bibinfo {author} {\bibfnamefont {V.}~\bibnamefont {Cherkez}},
  \bibinfo {author} {\bibfnamefont {C.}~\bibnamefont {Brun}}, \bibinfo {author}
  {\bibfnamefont {T.}~\bibnamefont {Cren}}, \bibinfo {author} {\bibfnamefont
  {C.}~\bibnamefont {Carbillet}}, \bibinfo {author} {\bibfnamefont
  {F.}~\bibnamefont {Debontridder}}, \bibinfo {author} {\bibfnamefont
  {K.}~\bibnamefont {Ilin}}, \bibinfo {author} {\bibfnamefont {M.}~\bibnamefont
  {Siegel}}, \bibinfo {author} {\bibfnamefont {A.}~\bibnamefont {Semenov}},
  \bibinfo {author} {\bibfnamefont {H.-W.}\ \bibnamefont {H\"ubers}},\ and\
  \bibinfo {author} {\bibfnamefont {D.}~\bibnamefont {Roditchev}},\ }\href@noop
  {} {\bibfield  {journal} {\bibinfo  {journal} {Phys. Rev. B}\ }\textbf
  {\bibinfo {volume} {88}},\ \bibinfo {pages} {014503} (\bibinfo {year}
  {2013})}\BibitemShut {NoStop}%
\bibitem [{\citenamefont {Dressel}(2013)}]{Dressel2013}%
  \BibitemOpen
  \bibfield  {author} {\bibinfo {author} {\bibfnamefont {M.}~\bibnamefont
  {Dressel}},\ }\href {https://doi.org/10.1155/2013/104379} {\bibfield
  {journal} {\bibinfo  {journal} {Adv. in Condens. Matt. Phys.}\ }\textbf
  {\bibinfo {volume} {2013}},\ \bibinfo {pages} {104379} (\bibinfo {year}
  {2013})}\BibitemShut {NoStop}%
\bibitem [{\citenamefont {\ifmmode~\check{S}\else \v{S}\fi{}indler}\ \emph
  {et~al.}(2018)\citenamefont {\ifmmode~\check{S}\else \v{S}\fi{}indler},
  \citenamefont {Kadlec}, \citenamefont {Ku\ifmmode~\check{z}\else
  \v{z}\fi{}el}, \citenamefont {Ilin}, \citenamefont {Siegel},\ and\
  \citenamefont {N\ifmmode~\check{e}\else \v{e}\fi{}mec}}]{Sindler2018}%
  \BibitemOpen
  \bibfield  {author} {\bibinfo {author} {\bibfnamefont {M.}~\bibnamefont
  {\ifmmode~\check{S}\else \v{S}\fi{}indler}}, \bibinfo {author} {\bibfnamefont
  {C.}~\bibnamefont {Kadlec}}, \bibinfo {author} {\bibfnamefont
  {P.}~\bibnamefont {Ku\ifmmode~\check{z}\else \v{z}\fi{}el}}, \bibinfo
  {author} {\bibfnamefont {K.}~\bibnamefont {Ilin}}, \bibinfo {author}
  {\bibfnamefont {M.}~\bibnamefont {Siegel}},\ and\ \bibinfo {author}
  {\bibfnamefont {H.}~\bibnamefont {N\ifmmode~\check{e}\else \v{e}\fi{}mec}},\
  }\href@noop {} {\bibfield  {journal} {\bibinfo  {journal} {Phys. Rev. B}\
  }\textbf {\bibinfo {volume} {97}},\ \bibinfo {pages} {054507} (\bibinfo
  {year} {2018})}\BibitemShut {NoStop}%
\bibitem [{\citenamefont {Henrich}\ \emph {et~al.}(2012)\citenamefont
  {Henrich}, \citenamefont {D\"orner}, \citenamefont {Hofherr}, \citenamefont
  {Il'in}, \citenamefont {Semenov}, \citenamefont {Heintze}, \citenamefont
  {Scheffler}, \citenamefont {Dressel},\ and\ \citenamefont
  {Siegel}}]{Henrich2012}%
  \BibitemOpen
  \bibfield  {author} {\bibinfo {author} {\bibfnamefont {D.}~\bibnamefont
  {Henrich}}, \bibinfo {author} {\bibfnamefont {S.}~\bibnamefont {D\"orner}},
  \bibinfo {author} {\bibfnamefont {M.}~\bibnamefont {Hofherr}}, \bibinfo
  {author} {\bibfnamefont {K.}~\bibnamefont {Il'in}}, \bibinfo {author}
  {\bibfnamefont {A.}~\bibnamefont {Semenov}}, \bibinfo {author} {\bibfnamefont
  {E.}~\bibnamefont {Heintze}}, \bibinfo {author} {\bibfnamefont
  {M.}~\bibnamefont {Scheffler}}, \bibinfo {author} {\bibfnamefont
  {M.}~\bibnamefont {Dressel}},\ and\ \bibinfo {author} {\bibfnamefont
  {M.}~\bibnamefont {Siegel}},\ }\href@noop {} {\bibfield  {journal} {\bibinfo
  {journal} {J. Appl. Phys.}\ }\textbf {\bibinfo {volume} {112}},\ \bibinfo
  {pages} {074511} (\bibinfo {year} {2012})}\BibitemShut {NoStop}%
\bibitem [{\citenamefont {Xi}(2011)}]{Xithesis}%
  \BibitemOpen
  \bibfield  {author} {\bibinfo {author} {\bibfnamefont {X.}~\bibnamefont
  {Xi}},\ }\emph {\bibinfo {title} {Conventional and time-resolved spectroscopy
  of magnetic properties of superconducting thin films}},\ \href@noop {}
  {\bibinfo {type} {dissertation thesis}},\ \bibinfo  {school} {University of
  Florida} (\bibinfo {year} {2011})\BibitemShut {NoStop}%
\bibitem [{\citenamefont {Poole~Jr.}\ \emph {et~al.}(2007)\citenamefont
  {Poole~Jr.}, \citenamefont {Farach}, \citenamefont {Creswick},\ and\
  \citenamefont {Prozorov}}]{Poole}%
  \BibitemOpen
  \bibfield  {author} {\bibinfo {author} {\bibfnamefont {C.~P.}\ \bibnamefont
  {Poole~Jr.}}, \bibinfo {author} {\bibfnamefont {H.~A.}\ \bibnamefont
  {Farach}}, \bibinfo {author} {\bibfnamefont {R.~J.}\ \bibnamefont
  {Creswick}},\ and\ \bibinfo {author} {\bibfnamefont {R.}~\bibnamefont
  {Prozorov}},\ }\href@noop {} {\emph {\bibinfo {title} {Superconductivity}}}\
  (\bibinfo  {publisher} {Academic Press},\ \bibinfo {year} {2007})\BibitemShut
  {NoStop}%
\bibitem [{\citenamefont {Nahata}, \citenamefont {Yardley},\ and\ \citenamefont
  {Heinz}(1999)}]{Nahata1999}%
  \BibitemOpen
  \bibfield  {author} {\bibinfo {author} {\bibfnamefont {A.}~\bibnamefont
  {Nahata}}, \bibinfo {author} {\bibfnamefont {J.~T.}\ \bibnamefont
  {Yardley}},\ and\ \bibinfo {author} {\bibfnamefont {T.~F.}\ \bibnamefont
  {Heinz}},\ }\href@noop {} {\bibfield  {journal} {\bibinfo  {journal} {Appl.
  Phys. Lett.}\ }\textbf {\bibinfo {volume} {75}},\ \bibinfo {pages} {2524}
  (\bibinfo {year} {1999})}\BibitemShut {NoStop}%
\bibitem [{poz()}]{poznamka}%
  \BibitemOpen
  \href@noop {} {}\bibinfo {note} {We are using the convention $E(t)=E_0 \exp
  (-i\omega t)$ which implies the following forms of the complex quantities:
  conductivity $\tilde{\sigma}=\sigma_1+i \sigma_2$, permittivity
  $\tilde{\varepsilon}=\varepsilon_1+i\varepsilon_2$, and refractive index
  $\tilde{n}=n+i\kappa$.}\BibitemShut {Stop}%
\bibitem [{\citenamefont {Alsthuler}\ and\ \citenamefont
  {Aronov}(1987)}]{Altshuler1987}%
  \BibitemOpen
  \bibfield  {author} {\bibinfo {author} {\bibfnamefont {B.}~\bibnamefont
  {Alsthuler}}\ and\ \bibinfo {author} {\bibfnamefont {A.}~\bibnamefont
  {Aronov}},\ }\enquote {\bibinfo {title} {Electron-electron interactions in
  disordered systems},}\ \ (\bibinfo  {publisher} {North-Holland},\ \bibinfo
  {address} {Amsterdam},\ \bibinfo {year} {1987})\BibitemShut {NoStop}%
\bibitem [{\citenamefont {Neilinger}\ \emph {et~al.}(2019)\citenamefont
  {Neilinger}, \citenamefont {Gregu\ifmmode~\check{s}\else \v{s}\fi{}},
  \citenamefont {Manca}, \citenamefont {Gran\ifmmode \check{c}\else
  \v{c}\fi{}i\ifmmode~\check{c}\else \v{c}\fi{}}, \citenamefont
  {Kop\ifmmode~\check{c}\else \v{c}\fi{}\'{\i}k}, \citenamefont {Szab\'o},
  \citenamefont {Samuely}, \citenamefont {Hlubina},\ and\ \citenamefont
  {Grajcar}}]{Neilinger2019}%
  \BibitemOpen
  \bibfield  {author} {\bibinfo {author} {\bibfnamefont {P.}~\bibnamefont
  {Neilinger}}, \bibinfo {author} {\bibfnamefont {J.}~\bibnamefont
  {Gregu\ifmmode~\check{s}\else \v{s}\fi{}}}, \bibinfo {author} {\bibfnamefont
  {D.}~\bibnamefont {Manca}}, \bibinfo {author} {\bibfnamefont
  {B.}~\bibnamefont {Gran\ifmmode \check{c}\else
  \v{c}\fi{}i\ifmmode~\check{c}\else \v{c}\fi{}}}, \bibinfo {author}
  {\bibfnamefont {M.}~\bibnamefont {Kop\ifmmode~\check{c}\else
  \v{c}\fi{}\'{\i}k}}, \bibinfo {author} {\bibfnamefont {P.}~\bibnamefont
  {Szab\'o}}, \bibinfo {author} {\bibfnamefont {P.}~\bibnamefont {Samuely}},
  \bibinfo {author} {\bibfnamefont {R.}~\bibnamefont {Hlubina}},\ and\ \bibinfo
  {author} {\bibfnamefont {M.}~\bibnamefont {Grajcar}},\ }\href@noop {}
  {\bibfield  {journal} {\bibinfo  {journal} {Phys. Rev. B}\ }\textbf {\bibinfo
  {volume} {100}},\ \bibinfo {pages} {241106} (\bibinfo {year}
  {2019})}\BibitemShut {NoStop}%
\bibitem [{\citenamefont {Cheng}\ \emph {et~al.}(2016)\citenamefont {Cheng},
  \citenamefont {Wu}, \citenamefont {Laurita}, \citenamefont {Singh},
  \citenamefont {Chand}, \citenamefont {Raychaudhuri},\ and\ \citenamefont
  {Armitage}}]{Cheng2016}%
  \BibitemOpen
  \bibfield  {author} {\bibinfo {author} {\bibfnamefont {B.}~\bibnamefont
  {Cheng}}, \bibinfo {author} {\bibfnamefont {L.}~\bibnamefont {Wu}}, \bibinfo
  {author} {\bibfnamefont {N.~J.}\ \bibnamefont {Laurita}}, \bibinfo {author}
  {\bibfnamefont {H.}~\bibnamefont {Singh}}, \bibinfo {author} {\bibfnamefont
  {M.}~\bibnamefont {Chand}}, \bibinfo {author} {\bibfnamefont
  {P.}~\bibnamefont {Raychaudhuri}},\ and\ \bibinfo {author} {\bibfnamefont
  {N.~P.}\ \bibnamefont {Armitage}},\ }\href@noop {} {\bibfield  {journal}
  {\bibinfo  {journal} {Phys. Rev. B}\ }\textbf {\bibinfo {volume} {93}},\
  \bibinfo {pages} {180511} (\bibinfo {year} {2016})}\BibitemShut {NoStop}%
\bibitem [{Sup()}]{SupMat}%
  \BibitemOpen
  \href@noop {} {}\bibinfo {note} {See Supplemental Material at
  -URL-}\BibitemShut {NoStop}%
\bibitem [{\citenamefont {Herman}\ and\ \citenamefont
  {Hlubina}(2017)}]{Herman2017}%
  \BibitemOpen
  \bibfield  {author} {\bibinfo {author} {\bibfnamefont {F.}~\bibnamefont
  {Herman}}\ and\ \bibinfo {author} {\bibfnamefont {R.}~\bibnamefont
  {Hlubina}},\ }\href@noop {} {\bibfield  {journal} {\bibinfo  {journal} {Phys.
  Rev. B}\ }\textbf {\bibinfo {volume} {96}},\ \bibinfo {pages} {014509}
  (\bibinfo {year} {2017})}\BibitemShut {NoStop}%
\bibitem [{\citenamefont {Dynes}, \citenamefont {Narayanamurti},\ and\
  \citenamefont {Garno}(1978)}]{Dynes1978}%
  \BibitemOpen
  \bibfield  {author} {\bibinfo {author} {\bibfnamefont {R.~C.}\ \bibnamefont
  {Dynes}}, \bibinfo {author} {\bibfnamefont {V.}~\bibnamefont
  {Narayanamurti}},\ and\ \bibinfo {author} {\bibfnamefont {J.~P.}\
  \bibnamefont {Garno}},\ }\href {https://doi.org/10.1103/PhysRevLett.41.1509}
  {\bibfield  {journal} {\bibinfo  {journal} {Phys. Rev. Lett.}\ }\textbf
  {\bibinfo {volume} {41}},\ \bibinfo {pages} {1509} (\bibinfo {year}
  {1978})}\BibitemShut {NoStop}%
\bibitem [{\citenamefont {Zimmermann}\ \emph {et~al.}(1991)\citenamefont
  {Zimmermann}, \citenamefont {Brandt}, \citenamefont {Bauer}, \citenamefont
  {Seider},\ and\ \citenamefont {Genzel}}]{Zimmermann1991}%
  \BibitemOpen
  \bibfield  {author} {\bibinfo {author} {\bibfnamefont {W.}~\bibnamefont
  {Zimmermann}}, \bibinfo {author} {\bibfnamefont {E.}~\bibnamefont {Brandt}},
  \bibinfo {author} {\bibfnamefont {M.}~\bibnamefont {Bauer}}, \bibinfo
  {author} {\bibfnamefont {E.}~\bibnamefont {Seider}},\ and\ \bibinfo {author}
  {\bibfnamefont {L.}~\bibnamefont {Genzel}},\ }\href@noop {} {\bibfield
  {journal} {\bibinfo  {journal} {Physica C: Superconductivity}\ }\textbf
  {\bibinfo {volume} {183}},\ \bibinfo {pages} {99} (\bibinfo {year}
  {1991})}\BibitemShut {NoStop}%
\bibitem [{THz()}]{THz}%
  \BibitemOpen
  \href@noop {} {}\bibinfo {note} {Throughout the article we express energy in
  THz units as it allows for a direct comparison with experimental spectra
  $\sigma(\nu)$. The energy can be calculated by multiplying by Planck constant
  $h$. The scattering time $\tau$ is related to the energy scale $E$ by
  $E=\hbar / \tau$ where $\hbar$ is the reduced Planck constant.}\BibitemShut
  {Stop}%
\bibitem [{\citenamefont {\v{S}indler}\ \emph {et~al.}(2017)\citenamefont
  {\v{S}indler}, \citenamefont {Tesa\v{r}}, \citenamefont {Kol\'a\v{c}ek},\
  and\ \citenamefont {Skrbek}}]{Sindler2017}%
  \BibitemOpen
  \bibfield  {author} {\bibinfo {author} {\bibfnamefont {M.}~\bibnamefont
  {\v{S}indler}}, \bibinfo {author} {\bibfnamefont {R.}~\bibnamefont
  {Tesa\v{r}}}, \bibinfo {author} {\bibfnamefont {J.}~\bibnamefont
  {Kol\'a\v{c}ek}},\ and\ \bibinfo {author} {\bibfnamefont {L.}~\bibnamefont
  {Skrbek}},\ }\href@noop {} {\bibfield  {journal} {\bibinfo  {journal}
  {Physica C: Superconductivity its Applications}\ }\textbf {\bibinfo {volume}
  {533}},\ \bibinfo {pages} {154} (\bibinfo {year} {2017})},\ \bibinfo {note}
  {ninth international conference on Vortex Matter in nanostructured
  Superdonductors}\BibitemShut {NoStop}%
\bibitem [{\citenamefont {Luzhbin}(2001)}]{Luzhbin2001}%
  \BibitemOpen
  \bibfield  {author} {\bibinfo {author} {\bibfnamefont {D.~A.}\ \bibnamefont
  {Luzhbin}},\ }\href@noop {} {\bibfield  {journal} {\bibinfo  {journal} {Phys.
  Solid State}\ }\textbf {\bibinfo {volume} {43}},\ \bibinfo {pages} {1823}
  (\bibinfo {year} {2001})}\BibitemShut {NoStop}%
\bibitem [{\citenamefont {Xi}\ \emph {et~al.}(2010)\citenamefont {Xi},
  \citenamefont {Hwang}, \citenamefont {Martin}, \citenamefont {Tanner},\ and\
  \citenamefont {Carr}}]{Xi2010}%
  \BibitemOpen
  \bibfield  {author} {\bibinfo {author} {\bibfnamefont {X.}~\bibnamefont
  {Xi}}, \bibinfo {author} {\bibfnamefont {J.}~\bibnamefont {Hwang}}, \bibinfo
  {author} {\bibfnamefont {C.}~\bibnamefont {Martin}}, \bibinfo {author}
  {\bibfnamefont {D.~B.}\ \bibnamefont {Tanner}},\ and\ \bibinfo {author}
  {\bibfnamefont {G.~L.}\ \bibnamefont {Carr}},\ }\href@noop {} {\bibfield
  {journal} {\bibinfo  {journal} {Phys. Rev. Lett.}\ }\textbf {\bibinfo
  {volume} {105}},\ \bibinfo {pages} {257006} (\bibinfo {year}
  {2010})}\BibitemShut {NoStop}%
\bibitem [{\citenamefont {Skalski}, \citenamefont {Betbeder-Matibet},\ and\
  \citenamefont {Weiss}(1964)}]{Skalski1964}%
  \BibitemOpen
  \bibfield  {author} {\bibinfo {author} {\bibfnamefont {S.}~\bibnamefont
  {Skalski}}, \bibinfo {author} {\bibfnamefont {O.}~\bibnamefont
  {Betbeder-Matibet}},\ and\ \bibinfo {author} {\bibfnamefont {P.~R.}\
  \bibnamefont {Weiss}},\ }\href@noop {} {\bibfield  {journal} {\bibinfo
  {journal} {Phys. Rev.}\ }\textbf {\bibinfo {volume} {136}},\ \bibinfo {pages}
  {A1500} (\bibinfo {year} {1964})}\BibitemShut {NoStop}%
\bibitem [{\citenamefont {Garnett}\ and\ \citenamefont
  {Larmor}(1904)}]{Garnett1904}%
  \BibitemOpen
  \bibfield  {author} {\bibinfo {author} {\bibfnamefont {J.~C.~M.}\
  \bibnamefont {Garnett}}\ and\ \bibinfo {author} {\bibfnamefont
  {J.}~\bibnamefont {Larmor}},\ }\href@noop {} {\bibfield  {journal} {\bibinfo
  {journal} {Philos. Trans. Royal Soc. of London. Series A, Containing Papers
  of a Mathematical or Phys. Character}\ }\textbf {\bibinfo {volume} {203}},\
  \bibinfo {pages} {385} (\bibinfo {year} {1904})}\BibitemShut {NoStop}%
\bibitem [{\citenamefont {Rychetsk\'y}, \citenamefont {Glogarov\'a},\ and\
  \citenamefont {Novotn\'a}(2004)}]{Rychetsky2004}%
  \BibitemOpen
  \bibfield  {author} {\bibinfo {author} {\bibfnamefont {I.}~\bibnamefont
  {Rychetsk\'y}}, \bibinfo {author} {\bibfnamefont {M.}~\bibnamefont
  {Glogarov\'a}},\ and\ \bibinfo {author} {\bibfnamefont {V.}~\bibnamefont
  {Novotn\'a}},\ }\href@noop {} {\bibfield  {journal} {\bibinfo  {journal}
  {Ferroelectrics}\ }\textbf {\bibinfo {volume} {300}},\ \bibinfo {pages} {135}
  (\bibinfo {year} {2004})}\BibitemShut {NoStop}%
\bibitem [{\citenamefont {Tinkham}(1996)}]{Tinkham1996}%
  \BibitemOpen
  \bibfield  {author} {\bibinfo {author} {\bibfnamefont {M.}~\bibnamefont
  {Tinkham}},\ }\href@noop {} {\emph {\bibinfo {title} {Introduction to
  Superconductivity}}}\ (\bibinfo  {publisher} {McGraw-Hill},\ \bibinfo
  {address} {New York},\ \bibinfo {year} {1996})\BibitemShut {NoStop}%
\bibitem [{\citenamefont {Fulde}(2010)}]{Fulde2010}%
  \BibitemOpen
  \bibfield  {author} {\bibinfo {author} {\bibfnamefont {P.}~\bibnamefont
  {Fulde}},\ }\href@noop {} {\bibfield  {journal} {\bibinfo  {journal} {Modern
  Phys. Lett. B}\ }\textbf {\bibinfo {volume} {24}},\ \bibinfo {pages} {2601}
  (\bibinfo {year} {2010})}\BibitemShut {NoStop}%
\bibitem [{\citenamefont {Xi}\ \emph {et~al.}(2013)\citenamefont {Xi},
  \citenamefont {Park}, \citenamefont {Graf}, \citenamefont {Carr},\ and\
  \citenamefont {Tanner}}]{Xi2013}%
  \BibitemOpen
  \bibfield  {author} {\bibinfo {author} {\bibfnamefont {X.}~\bibnamefont
  {Xi}}, \bibinfo {author} {\bibfnamefont {J.-H.}\ \bibnamefont {Park}},
  \bibinfo {author} {\bibfnamefont {D.}~\bibnamefont {Graf}}, \bibinfo {author}
  {\bibfnamefont {G.~L.}\ \bibnamefont {Carr}},\ and\ \bibinfo {author}
  {\bibfnamefont {D.~B.}\ \bibnamefont {Tanner}},\ }\href
  {https://doi.org/10.1103/PhysRevB.87.184503} {\bibfield  {journal} {\bibinfo
  {journal} {Phys. Rev. B}\ }\textbf {\bibinfo {volume} {87}},\ \bibinfo
  {pages} {184503} (\bibinfo {year} {2013})}\BibitemShut {NoStop}%
\bibitem [{\citenamefont {Sheahen}(1966)}]{Sheahen1966}%
  \BibitemOpen
  \bibfield  {author} {\bibinfo {author} {\bibfnamefont {T.~P.}\ \bibnamefont
  {Sheahen}},\ }\href@noop {} {\bibfield  {journal} {\bibinfo  {journal} {Phys.
  Rev.}\ }\textbf {\bibinfo {volume} {149}},\ \bibinfo {pages} {368} (\bibinfo
  {year} {1966})}\BibitemShut {NoStop}%
\bibitem [{\citenamefont {{\v{S}}indler}\ \emph {et~al.}(2014)\citenamefont
  {{\v{S}}indler}, \citenamefont {Tesa{\v{r}}}, \citenamefont
  {Kol{\'{a}}{\v{c}}ek}, \citenamefont {Szab{\'{o}}}, \citenamefont {Samuely},
  \citenamefont {Ha{\v{s}}kov{\'{a}}}, \citenamefont {Kadlec}, \citenamefont
  {Kadlec},\ and\ \citenamefont {Ku{\v{z}}el}}]{Sindler2014}%
  \BibitemOpen
  \bibfield  {author} {\bibinfo {author} {\bibfnamefont {M.}~\bibnamefont
  {{\v{S}}indler}}, \bibinfo {author} {\bibfnamefont {R.}~\bibnamefont
  {Tesa{\v{r}}}}, \bibinfo {author} {\bibfnamefont {J.}~\bibnamefont
  {Kol{\'{a}}{\v{c}}ek}}, \bibinfo {author} {\bibfnamefont {P.}~\bibnamefont
  {Szab{\'{o}}}}, \bibinfo {author} {\bibfnamefont {P.}~\bibnamefont
  {Samuely}}, \bibinfo {author} {\bibfnamefont {V.}~\bibnamefont
  {Ha{\v{s}}kov{\'{a}}}}, \bibinfo {author} {\bibfnamefont {C.}~\bibnamefont
  {Kadlec}}, \bibinfo {author} {\bibfnamefont {F.}~\bibnamefont {Kadlec}},\
  and\ \bibinfo {author} {\bibfnamefont {P.}~\bibnamefont {Ku{\v{z}}el}},\
  }\href@noop {} {\bibfield  {journal} {\bibinfo  {journal} {Superconductor
  Sci. Technol.}\ }\textbf {\bibinfo {volume} {27}},\ \bibinfo {pages} {055009}
  (\bibinfo {year} {2014})}\BibitemShut {NoStop}%
\bibitem [{\citenamefont {Szab\'o}\ \emph {et~al.}(2016)\citenamefont
  {Szab\'o}, \citenamefont {Samuely}, \citenamefont {Ha\ifmmode~\check{s}\else
  \v{s}\fi{}kov\'a}, \citenamefont {Ka\ifmmode \check{c}\else
  \v{c}\fi{}mar\ifmmode~\check{c}\else \v{c}\fi{}\'{\i}k}, \citenamefont
  {\ifmmode \check{Z}\else \v{Z}\fi{}emli\ifmmode~\check{c}\else \v{c}\fi{}ka},
  \citenamefont {Grajcar}, \citenamefont {Rodrigo},\ and\ \citenamefont
  {Samuely}}]{Szabo2016}%
  \BibitemOpen
  \bibfield  {author} {\bibinfo {author} {\bibfnamefont {P.}~\bibnamefont
  {Szab\'o}}, \bibinfo {author} {\bibfnamefont {T.}~\bibnamefont {Samuely}},
  \bibinfo {author} {\bibfnamefont {V.}~\bibnamefont {Ha\ifmmode~\check{s}\else
  \v{s}\fi{}kov\'a}}, \bibinfo {author} {\bibfnamefont {J.}~\bibnamefont
  {Ka\ifmmode \check{c}\else \v{c}\fi{}mar\ifmmode~\check{c}\else
  \v{c}\fi{}\'{\i}k}}, \bibinfo {author} {\bibfnamefont {M.}~\bibnamefont
  {\ifmmode \check{Z}\else \v{Z}\fi{}emli\ifmmode~\check{c}\else
  \v{c}\fi{}ka}}, \bibinfo {author} {\bibfnamefont {M.}~\bibnamefont
  {Grajcar}}, \bibinfo {author} {\bibfnamefont {J.~G.}\ \bibnamefont
  {Rodrigo}},\ and\ \bibinfo {author} {\bibfnamefont {P.}~\bibnamefont
  {Samuely}},\ }\href {https://doi.org/10.1103/PhysRevB.93.014505} {\bibfield
  {journal} {\bibinfo  {journal} {Phys. Rev. B}\ }\textbf {\bibinfo {volume}
  {93}},\ \bibinfo {pages} {014505} (\bibinfo {year} {2016})}\BibitemShut
  {NoStop}%
\end{thebibliography}%

\end{document}


\title{Supplemental material: Onset of superconductor-insulator transition in a thin NbN film under in-plane magnetic field studied by terahertz spectroscopy}

\author{M. \v{S}indler}
\author{F.~Kadlec}
\author{C.~Kadlec}
\affiliation{Institute of Physics, Academy of Sciences, 16253 Prague 6, Czech Republic}

\date{\today}

\maketitle
In the paper, we presented  our experimental data for either  $E^\|$ or $E^{\perp}$ linear polarisation which were evaluated either by using equation (1) or by equation (3) from the paper.
Here, we show the complete set of graphs which relates to figures 1-3 from the paper. For further details, see the paper text.

\begin{figure}[h]
\centering
\includegraphics[width=0.8\columnwidth]{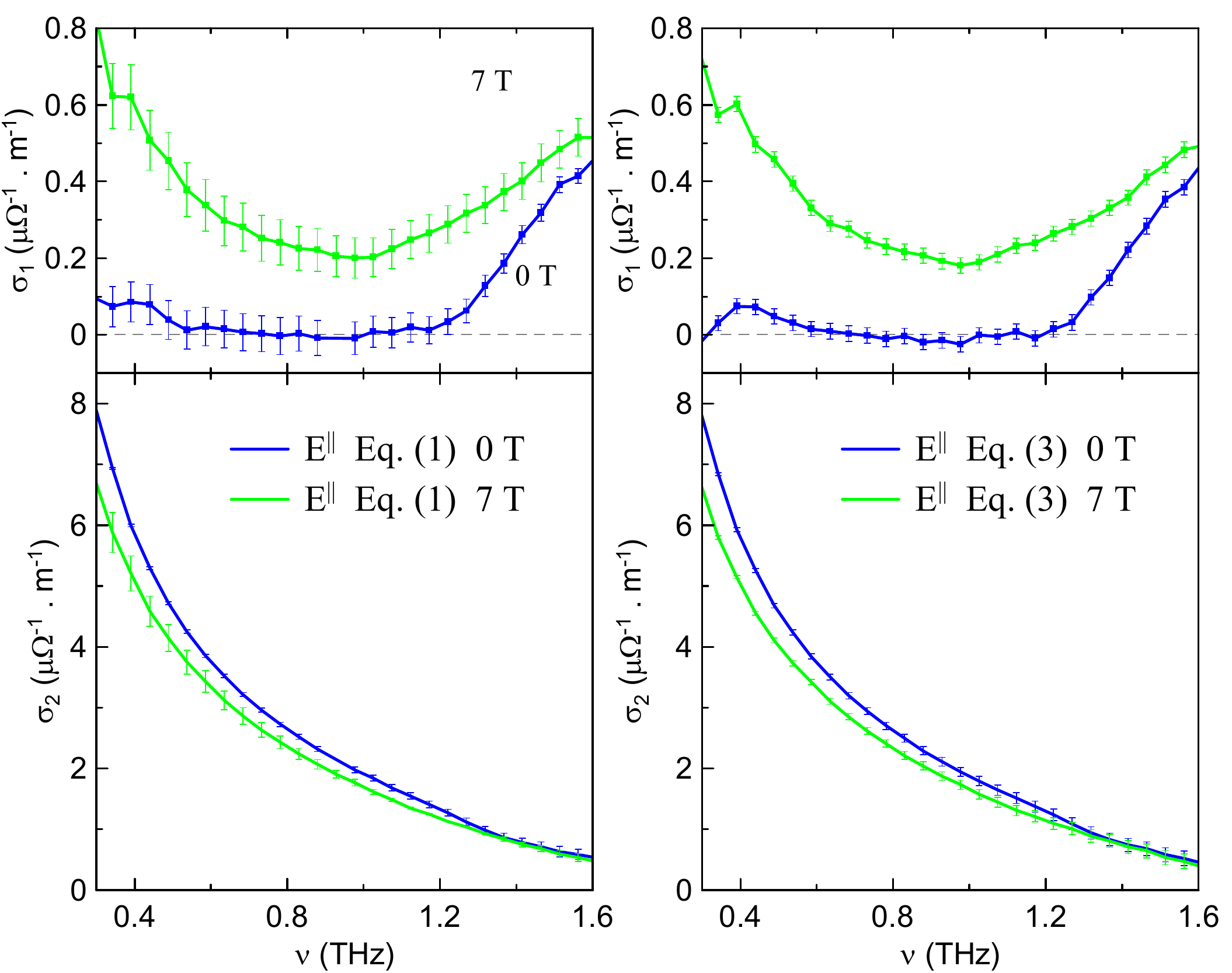}
\includegraphics[width=0.8 \columnwidth]{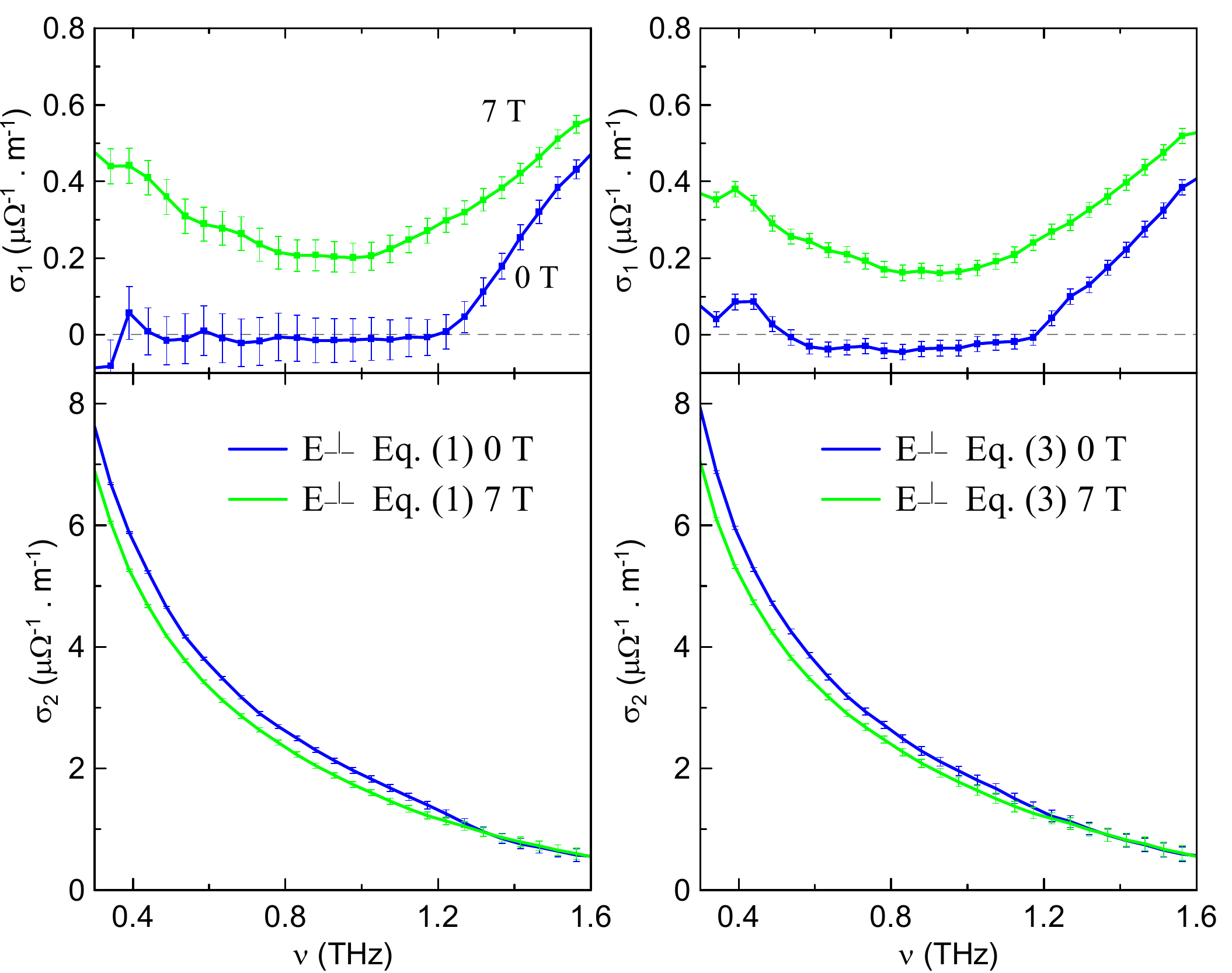}
\caption{
  Real (shown also in the paper) and imaginary part spectra of NbN conductivity $\sigma_1(\nu)$ together with estimated errorbars 
for $E^\|$ and $E^{\perp}$ obtained by  the two methods described in the paper text at $T=3\,\rm K$ and a magnetic field of 0 and 7\,T. Note that the evaluation of the
imaginary part $\sigma_2(\nu)$ provides spectra with a much lower uncertainty.
}
\label{fig:compare}
\end{figure}

\begin{figure}[h]
\centering
\includegraphics[width=0.8\columnwidth]{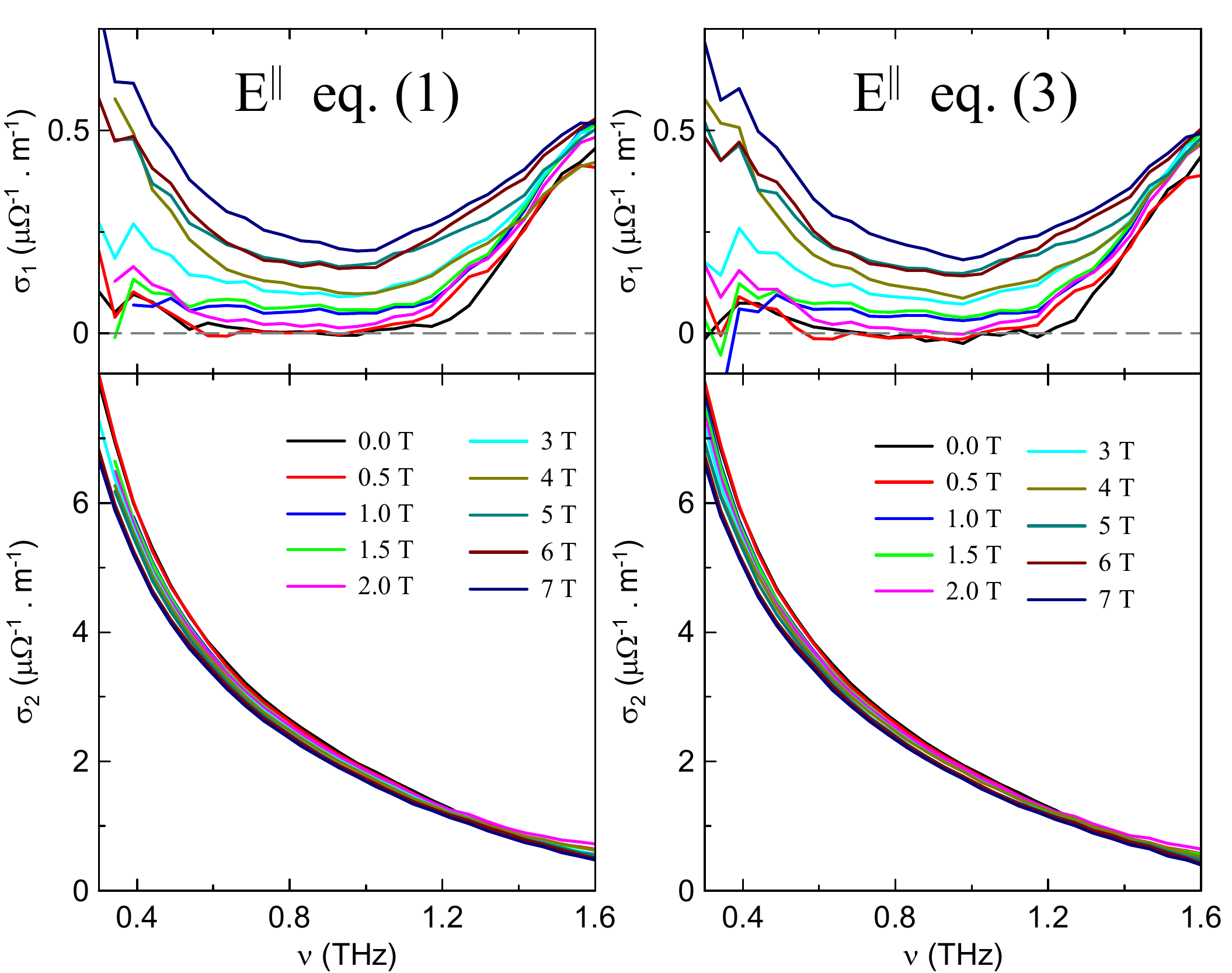}
\includegraphics[width=0.8\columnwidth]{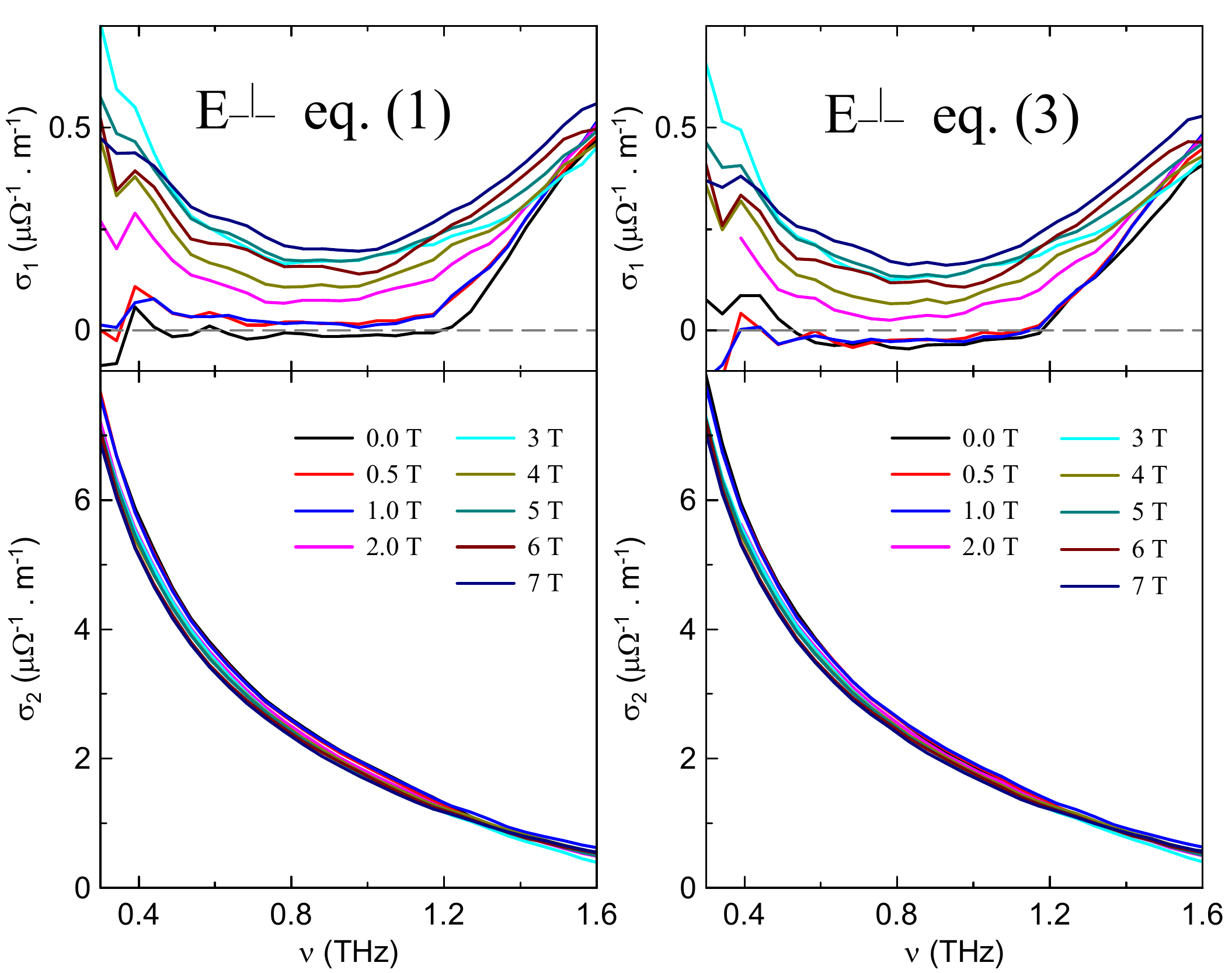}
\caption{
  Real  and imaginary part spectra of NbN conductivity $\sigma(\nu)$ obtained by
  the two described methods from experiments using both linear
polarizations $E^\|$ and $E^{\perp}$ at $T=3\,\rm K$ and a magnetic field from 0~T up to 7\,T. 
Experimental data for  $E^\|$ and $E^{\perp}$ are similar, except  $\sigma_1$ for 3\,T  and  $E^{\perp}$ polarisation which is an outlier deviating from the remaining experiments.
}
\label{fig:compare}
\end{figure}

\begin{figure}[h]
\centering
\includegraphics[width=0.8\columnwidth]{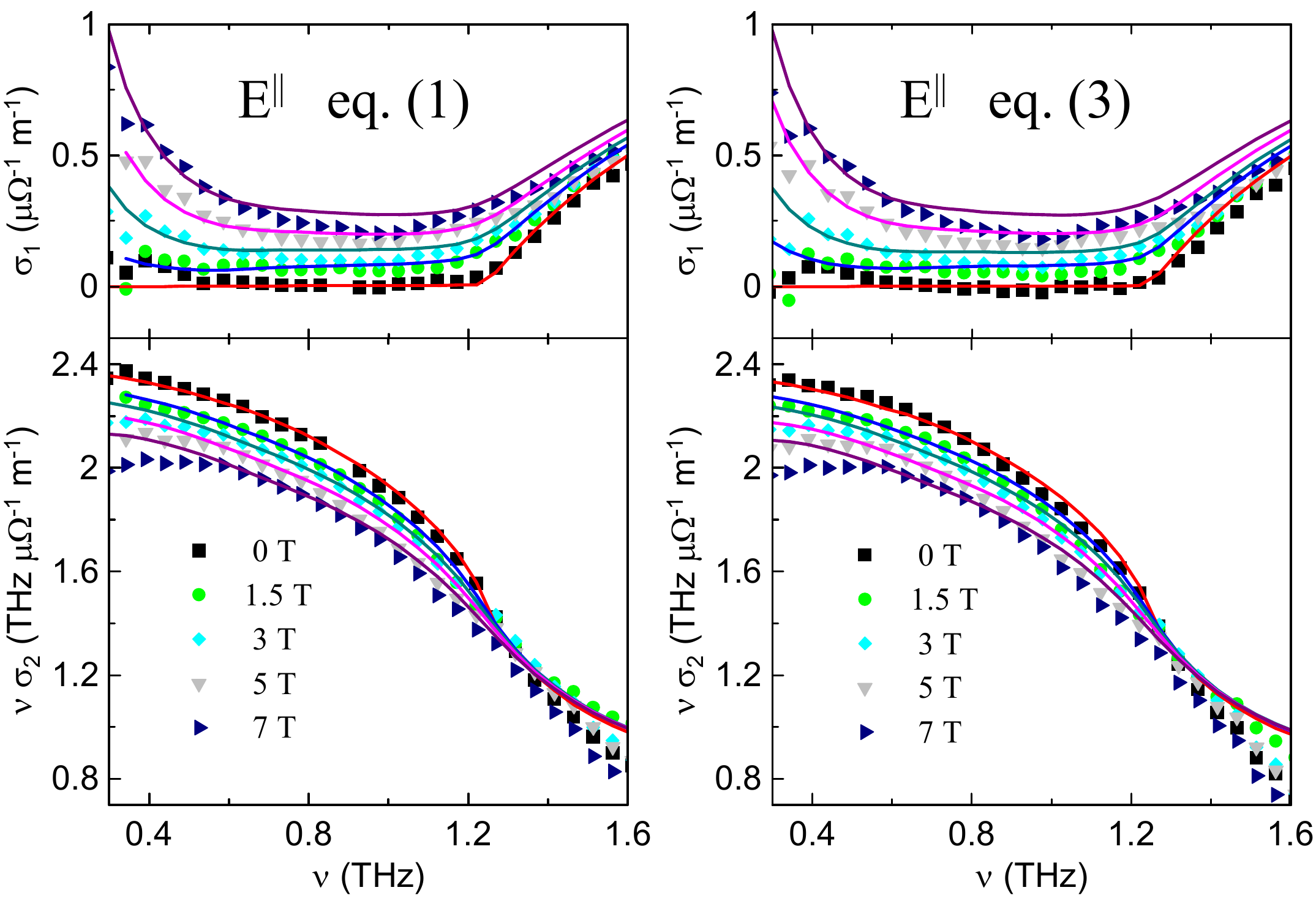}
\includegraphics[width=0.8\columnwidth]{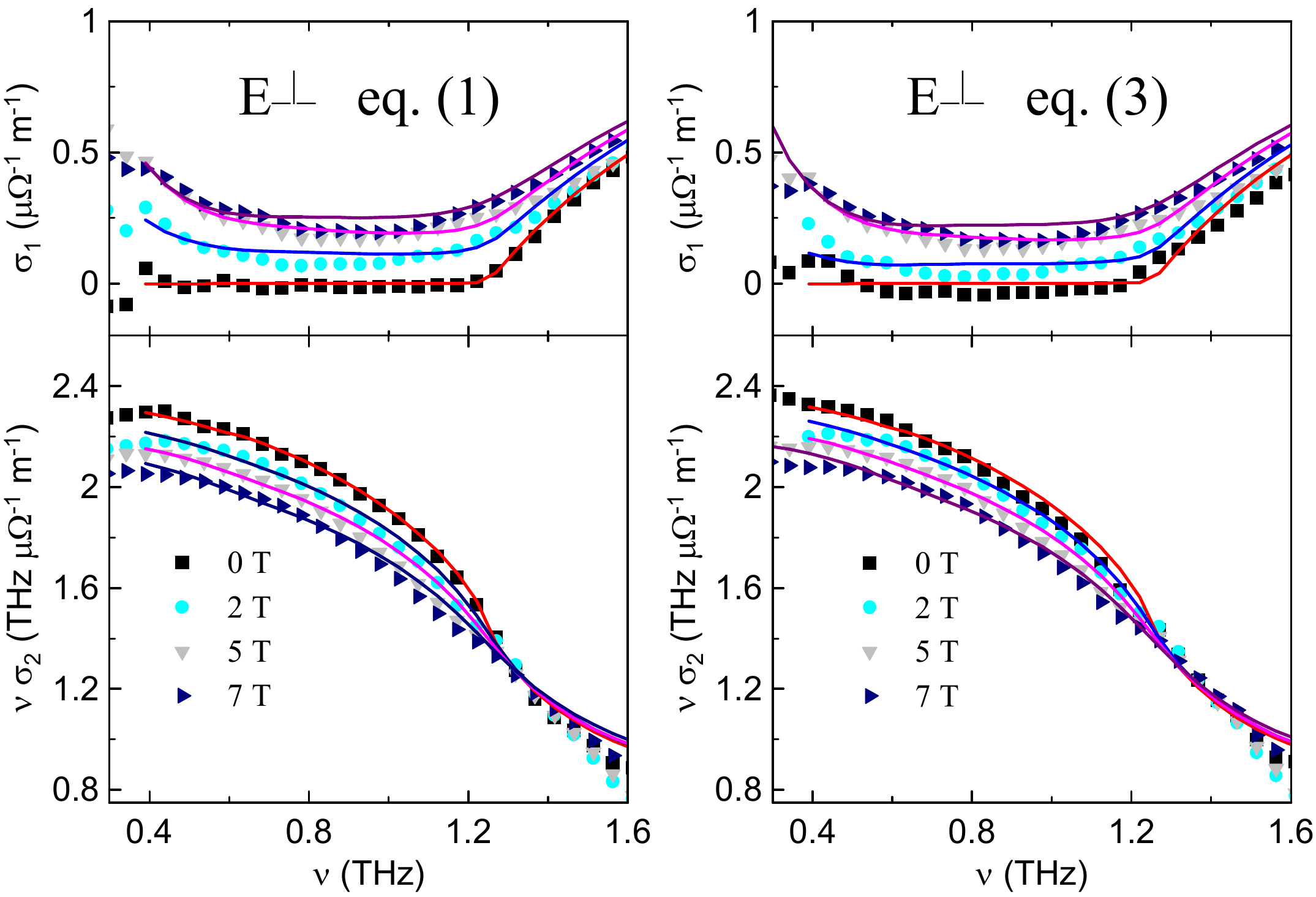}
\caption{
  Symbols: real and imaginary part of the conductivity of NbN at $T=3\,\rm K$. Lines are fits by the Maxwell-Garnett formula using 
  the Drude model for the normal-state
  $\tilde{\sigma}_{\rm n}(\nu)$ and the HH model for the
  superconducting-state $\tilde{\sigma}_{\rm s}(\nu)$ components,
  respectively. The imaginary part is plotted as $\nu
\sigma_2(\nu)$ for improved clarity. 
}

\label{fig:compare}
\end{figure}

\bibliographystyle{aipnum4-1}
\bibliography{NbN.bib}